\documentclass[aps,pra,a4paper,twocolumn,superscriptaddress,floatfix,longbibliography
]{revtex4-1}

\usepackage[utf8]{inputenc}
\usepackage[german,english]{babel}
\usepackage[colorlinks=true,allcolors=blue]{hyperref}

\usepackage{graphicx}
\usepackage{amsmath}
\usepackage{booktabs}
\usepackage{amsfonts}
\usepackage{amssymb}
\usepackage{dsfont}
\usepackage{color}
\usepackage{enumitem}
\setlist[description]{labelindent=8pt,style=multiline,leftmargin=1.8cm}

\newcommand{\PP}{\text{p}}
\newcommand{\DP}{\Delta \text{p}}
\newcommand{\E}{\mathds{E}}
\newcommand{\equal}{%
  \underset{{\raisebox{2pt}[0pt][0pt]{$\scriptstyle 2\pi i$}}}{=}}

\newcommand{\white}[1]{\begingroup\color{white}#1\endgroup}
\newcommand{\bra}[1]{\langle #1 |}
\newcommand{\ket}[1]{| #1 \rangle}
\newcommand{\norm}[1]{\lVert #1 \rVert}
\newcommand{\CZ}{\ensuremath{\mathcal{C}\text{-Z}}}
\newcommand{\one}{\mathds{1}}
\renewcommand{\Re}{\mathrm{Re}}

\begin{document}

\title{Estimation of coherent error sources from stabilizer measurements}

\author{Davide Orsucci}
\affiliation{Institut f{\"u}r Theoretische Physik,
Universit{\"a}t Innsbruck, Technikerstra{\ss }e 21 A, A-6020 Innsbruck}

\author{Markus Tiersch}
\affiliation{Institut f{\"u}r Theoretische Physik,
Universit{\"a}t Innsbruck, Technikerstra{\ss }e 21 A, A-6020 Innsbruck}

\author{Hans J.~Briegel}
\affiliation{Institut f{\"u}r Theoretische Physik,
Universit{\"a}t Innsbruck, Technikerstra{\ss }e 21 A, A-6020 Innsbruck}

\begin{abstract}

In the context of measurement-based quantum computation a way of maintaining the coherence of a \emph{graph state} is to measure its stabilizer operators. Aside from performing quantum error correction, it is possible to exploit the information gained from these measurements to characterize and then counteract a coherent source of errors; that is, to determine all the parameters of an \emph{error channel} that applies a fixed -- but unknown -- unitary operation to the physical qubits. Such a channel is generated, e.g., by local stray fields that act on the qubits. We study the case in which each qubit of a given \emph{graph state} may see a different error channel and we focus on channels given by a rotation on the Bloch sphere around either the $\hat{\boldsymbol{x}}$, $\hat{\boldsymbol{y}}$ or $\hat{\boldsymbol{z}}$ axis, for which analytical results can be given in a compact form. The possibility of reconstructing the channels at all qubits depends non-trivially on the topology of the graph state. We prove via perturbation methods that the reconstruction process is robust and supplement the analytic results with numerical evidence.

\end{abstract}

\maketitle

\section{Introduction} \label{sec:introduction}

In order to construct devices that reliably process quantum information it is necessary to characterize, measure and eventually remove all the known sources of detrimental noise~\cite{Moroder2013, Schwarz2013, Ladd2010}. It is however inevitable that a certain amount of disturbance will continue to affect the apparatus. Hence it usually will be indispensable to perform quantum error correction (QEC)~\cite{Gottesman1997, Bravyi1998, Lidar2013, Nigg2014, Martini2015} to reduce the error rate to a tolerable amount.

Robust and scalable architectures for quantum information processing require that their elementary building blocks are readily calibrated and easily operated. Better scalability with less manual intervention can be achieved by engineering devices which are capable of autonomously characterizing and correcting the error sources. Also, since these systems will likely become increasingly more integrated and miniaturized~\cite{Blatt2013, Horn2015, OBrien2015, Wang2015}, it will be difficult to have physical access to their building parts once the apparatus has been assembled. Therefore, it is desirable that these devices are designed and built so that it is possible to gather data while the machine is running, and thus, on-the-fly, achieve good control and feedback on the internal functioning.

In this paper we introduce a method that allows for an automated analysis of the presence of error sources and envision an \emph{agent}, connected to the quantum device, which performs classical statistical analysis of the data in order to learn which noise processes are occurring on all qubits. A related case study of adapting to a unitary channel acting on an isolated qubit is given in Ref.~\cite{Tiersch2015}. Furthermore, as we argued before, the device will necessarily be equipped with a QEC code to protect the logical subspace from noise. Hence we set out to investigate the following natural question: \textit{to what extent is it possible to characterize the error sources affecting the apparatus, having access to the syndrome measurements alone?} More specifically, we examine the scenario in which the agent collects the syndrome measurements and performs statistical analysis over them to detect and characterize a unitary component in the noise. Therefore we avoid the necessity of adding additional sensors to monitor the performance of the quantum processing device. This approach is independent of the physical implementation of the apparatus and it can be studied through a multitude of different methods, see e.g.\ Refs.~\cite{Combes2014, Martini2014, Wiebe2014, Ferrie2015, Omkar2015}.

Once the unitary component of the error has been characterized it is possible, at least in principle, to completely correct it through the application of a counter-unitary operation, which undoes the effect of this component of the error channel. As an alternative, we might allow the quantum computation to take place in a ``floating frame of reference'', in which the logical quantum basis change in time in order to compensate for the rotation created by the unitary error channel.

Hereafter we will study the information that can be recovered from the most restrictive type of QEC. We consider the extreme case of a QEC code in which only one state is protected, i.e., the logical subspace has dimension one. This QEC method can be adopted to produce the resource state for measurement-based quantum computation (MBQC)~\cite{RB2001,Raussendorf2003}. In MBQC particular classes of quantum states, e.g.\ the \emph{cluster states} \cite{Briegel2001}, are \emph{universal}, meaning that they can be employed to perform arbitrary quantum computations just by applying local measurements. The cluster states are a subset of a more general family of quantum states associated to undirected graphs, which are called \emph{graph states}. A method for obtaining any graph state consists in measuring a complete set of stabilizer operators (a.k.a.\ \emph{correlators}) and in applying the required corrections whenever an ``error'' is detected. The output of this procedure is a unique quantum state, i.e.\ the required graph state. Another setting in which this QEC can be employed is for measurement-based quantum repeaters~\cite{Wolfgang1999, Michael2012, Michael2015}, in which specific graph states are used as resources to perform entanglement purification and entanglement swapping to transmit quantum information over long distances.

The outline of the paper is the following. In Sec.~\ref{sec:notation} we give a brief introduction to graph states and set the notation; in Sec.~\ref{sec:error_model} we describe and motivate the error model we analyse in the paper; in Sec.~\ref{sec:reconstruction} we detail the effect of the unitary error channel over the stabilizer measurements statistics; then in Sec.~\ref{sec:promise_settings} we describe a special \emph{promise setting} in which the problem becomes solvable even with limited resources. The central part of the work begins in Sec.~\ref{sec:solving_methods}, where we provide a general method for reconstructing the error channels in the promise settings; then in Sec.~\ref{sec:particular_classes} we apply the general methods given in Sec.~\ref{sec:solving_methods} to some specific classes of graph states; finally in Sec.~\ref{sec:error_propagation} we show that our reconstruction methods are resilient to small deviations from the model. We conclude by giving in Sec.~\ref{sec:extension} a completely different approach which allows us to reconstruct the unitary error in complete generality, at the cost of requiring some extra resources.

\section{Preliminary concepts and notation} \label{sec:notation}

A \emph{graph state} $\ket{G}$ is quantum state associated with an \emph{undirected graph} $G=(V,E)$ having \emph{vertices} $V$ and \emph{edges} $E = \{ (a,b) \,|\, a,b \in V \}$, which connect pairs of vertices; $\ket{G}$ is defined as
\begin{equation}
  \ket{G} := \prod_{(a,b) \in E} \CZ_{a,b} \ket{+}^{\otimes N} \,,
\end{equation}
where $\ket{+} := (\ket{0} + \ket{1})/\sqrt{2}$, $N := |G|$ is the number of qubits of the graph state (each vertex represents a qubit) and $\CZ_{a,b}$ denotes the (symmetric) controlled-$Z$ operation between the qubits at vertex $a$ and $b$. The \emph{neighbours} of any vertex $a \in V$ are given by the vertices in the set $\mathcal{N}_a := \{b \in V \,|\, (a,b) \in E\}$. We also use the notation $\mathcal{N}_a' := \mathcal{N}_a \cup \{ a \}$.

We say that two graph states are \emph{locally unitarily equivalent} ($LU$-equivalent) if it is possible to obtain one graph state from the other by applying unitary operations acting on individual qubits.

A standard method that allows us to preserve the coherence of a graph state $\ket{G}$ consists in repeatedly measuring the \emph{correlator} operators
\begin{equation}\label{correlator}
  K_a := X_a \bigotimes_{b \in \mathcal{N}_a} Z_b \,,
\end{equation}
$\forall a \in V$. Here $X_a$, $Y_a$ and $Z_a$ denote the Pauli matrices acting on the qubit in vertex $a$; we also use the shorthand $\boldsymbol{S}_a := (X_a, Y_a, Z_a)$. All these correlators commute pairwise -- thus all the measurements can be performed simultaneously -- and form a complete set of observables.

We explicitly distinguish the \emph{correlator} operators $K_a$ from the random variables $\kappa_a$ which represent the \emph{realized measurement outcomes} of $K_a$. The variables $\kappa_a$ assume values $\kappa_a \in \{0,1\}$, corresponding respectively to the $\{+1,-1\}$ eigenvalues of $K_a$. The graph state $\ket{G}$ is the unique common eigenvector with eigenvalue equal to $+1$ for all the correlators; equivalently, this is the state one obtains when $\kappa_a = 0$, $\forall a \in V$. Hence, when the measurements outcomes of the correlators are given by $\{\kappa_a\}_{a \in V}$, the system is projected into the state 
\begin{equation}
  \ket{G'} = \prod_{a \in V} \big( Z_a \big)^{\kappa_a} \ket{G} \,,
\end{equation}
where, for any quantum gate $U$, $U^{\kappa_a} = \one$ when $\kappa_a = 0$ and $U^{\kappa_a} = U$ when $\kappa_a = 1$. The graph state $\ket{G'}$ is equivalent to $\ket{G}$ for the purpose of performing MBQC. The Pauli errors $Z_a$ can be accounted for by a redefinition of the logical encoding of the qubits: i.e., when $\kappa_a = 1$, the sign of the state $\ket{1}$ has to be reversed (that is, we make the association $\ket{0}_L \Leftrightarrow \ket{0}$ and $\ket{1}_L \Leftrightarrow - \ket{1}$). Thus we can assume, without loss of generality, that the graph state is re-prepared after the measurement of all the correlators in the standard graph state $\ket{G}$. As stated in the introduction, from a quantum error correction perspective these correlators provide a stabilizer code which protects only one word, the state $\ket{G}$. 

For a thorough introduction to these concepts, see Ref.~\cite{Hein2006journal}.

\section{Error model} \label{sec:error_model}

We consider the following model for a unitary error source that acts on a graph state. Suppose that our graph state is physically implemented with spin-$\frac{1}{2}$ particles, each representing a qubit with the spin degree of freedom \footnote{Spin-$\frac{1}{2}$ particles might be effectively implemented as neutral atoms trapped in optical lattices, or as ion in a confining potential. However these real systems are more complex than the model we study in this paper. For example, the qubits are usually encoded within two energy levels of the atom or ion, and not directly in the angular momentum degrees of freedom of the particle.}.
These particles carry an intrinsic magnetic dipole moment; hence they are susceptible to undergo Larmor precession when immersed in a magnetic field, with angular frequency equal to $\omega = \gamma B$, where $B$ is the magnetic field intensity and $\gamma$ is the gyro-magnetic ratio of the particle. However, as long as these stray fields are not subject to rapid and stochastic variations in time, it is possible to measure them and then apply a counter-unitary operation which undoes the unwanted precession. Equivalently, if the precession rate of each particle is known, we may let the logical encoding of the qubits become time-dependent, in order to compensate at the logical level the physical precession of the spins of the system. Notice that we focus on unitary processes since, if the error channel is given by a general master equation, the decoherence process can be characterized and mitigated, but not completely corrected.

In summary, we assume that our graph state $\ket{G}$ is affected by a space-varying static \emph{stray field} which has a different strength at each qubit of the state. We further assume that all measurements are performed at discrete time steps $\Delta t$, so that the action of the field on each qubit is given by a $SU(2)$ operation of the form $U_a = \text{exp}(-i \, \gamma_a \Delta t \, \boldsymbol{B}_a \cdot \boldsymbol{S}_a/2)$. We remark that the results we will obtain in this paper will not depend on the specific implementation of the physical qubits, nor on the physical model of the error source; our methods will be applicable whenever the error process acting on the qubits is dominated by a unitary local operation acting on the qubits in the system. Nonetheless, for concreteness we will always refer to the above specified physical error model. Consequently the word \emph{field} can be assumed to designate any channel acting on single qubits and which is given by a fixed, but unknown, unitary mapping.

Notice that in the special case where the stray field is almost uniform and acts equally on all the qubits in the graph state, then an efficient way to achieve noise resilience is given by encoding the information in decoherence-free subspaces. The properties of these subspaces have been extensively studied elsewhere (see for a review Refs.~\cite{Lidar1998, Lidar2003}) and thus we will not discuss nor employ them in our work. 

In conclusion, at each time step -- and on each vertex of the graph -- the qubits are affected by the following local operations:
\begin{equation} \label{fields}
  U_a := e^{-i \lambda_a \hat{\boldsymbol{n}}_a \cdot \boldsymbol{S}_a/2} \,,
\end{equation}
where $a \in V$ denotes the site at which the field acts, $\lambda_a \in \mathds{R}$ is proportional to the strengths of the field (in our physical model $\lambda_a = \gamma \Delta t \,|\boldsymbol{B}_a|$) and $\hat{\boldsymbol{n}}_a = (n_{x}^a, n_{y}^a, n_{z}^a)$ (with $n_x^2 + n_y^2 + n_z^2 = 1$) is the normalized vector in $\mathds{R}^3$ that gives the axis around which the rotation is performed. Observe that the relation $n_x^2 + n_y^2 + n_z^2 = 1$ allows us to recover $n_y$ from $(n_x, n_z)$, assuming that its sign is always $+1$, since this can be enforced by the transformation $(\lambda \mapsto -\lambda,\hat{\boldsymbol{n}} \mapsto - \hat{\boldsymbol{n}})$, which is the identity on $SO(3)$.

Naturally, in any realistic scenario there will be also other uncharted error sources that act on the physical system, causing some decoherence in the system. This further noise will in general render the reconstruction of the fields more difficult, but it will not qualitatively affect our results.

\section{Effect of the unitary error on stabilizer measurements} \label{sec:reconstruction}

In our setting, we are considering a QEC scheme which aims to preserve the coherence of the graph state by performing the necessary correlator measurements and reading out the corresponding results. Thus, the whole process can be schematized as follows: the graph state is prepared in the standard reference state $\ket{G}$; we wait for a fixed time $\Delta t$, during which the error channels act on the physical qubits; then we measure the correlators, which project the graph state on some state $\ket{G'}$, which depends on the measurement outcomes of the correlators (i.e.\ on the \emph{measured syndromes}); we finally use the syndromes to bring back $\ket{G'}$ into the reference state $\ket{G}$, either physically or by changing the logical encoding of the physical qubits. Implicitly, we are assuming that the state preparation, correlator readout and the final correction can be done very fast, w.r.t.\ the waiting time $\Delta t$, and almost free of errors. Moreover we will assume for the moment that the stray field is the only error source present, i.e., no other error channel or decoherence process is acting on the qubits. We will lift this assumption in Sec.~\ref{sec:full_error_analysis}, where we will consider the effect of having a decoherence channel acting alongside the stray field.

As a first step in our analysis, we need to find the probability that, for each vertex $a \in V$, the measurement outcome $\kappa_a$ of the correlator $K_a$ is equal to $0$ or $1$, given that the graph state was initialized in a state $\ket{G}$ corresponding to a graph $G = (V,E)$, and assuming that a unitary $U_a$ has acted on each vertex. The result can be found via a straightforward computation, given in Appendix~\ref{app:error_probability}. We define the \emph{probability difference} $\DP_a \in [-1,+1]$ as the difference between the probability of getting a ``correct'' outcome ($\kappa_a = 0$) and the probability of getting an ``error detected'' outcome ($\kappa_a = 1$)
\begin{equation} \label{prob_diff_def}
  \DP_a :=  \, \PP (\kappa_a = 0) \,-\, \PP (\kappa_a = +1) \,;
\end{equation}
then we obtain the following formula for $\DP_a$:
\begin{align} \label{general}
  \DP_a =  
    & \ \ \Big( (n_x^a)^2 + \beta_a \big( 1 - (n_x^a)^2 \big) \Big) \ \times \nonumber\\
    & \times \prod_{b \in \mathcal{N}_a}
      \Big( (n_z^b)^2 + \beta_b \big( 1 - (n_z^b)^2 \big) \Big) \,,
\end{align}
where $\beta_a := \cos \lambda_a$ and $(\lambda_a, n_x^a, n_z^b)$ are the parameters that define the action of the field on vertex $a$ as given by Eq.~\eqref{fields}. 

Notice that from the probability differences alone it is impossible to reconstruct the correct sign of $n_{x}^a$ and $n_{z}^a$, as Eq.~\eqref{general} depends on these only through $(n_{x}^a)^2, (n_{z}^a)^2$, nor the correct value of $\lambda_a$ in the set $\{\pm \lambda_a \! + \! 2k\pi \,|\, k \in \mathds{Z}\}$, since we can only access $\beta_a = \cos \lambda_a$.

In principle, these ambiguities can be easily resolved. For example, it is sufficient to vary the delay time $\Delta t$ to linearly vary the parameters $\lambda_a$, which can then be discriminated through the functions $\cos \lambda_a$. However, these methods rely on performing some extra operations, not included in the basic stabilizer code, and therefore we will not consider them in the following. In Sec.~\ref{sec:repeaters} we will exploit another resource, which is automatically included in MBQC, in order to reconstruct the fields. That is, we will use that some graph states are $LU$-equivalent even if they correspond to different graphs. $LU$-equivalent graph states have the same power for performing MBQC, assuming that arbitrary local measurements can be performed on the qubits; nonetheless the information that can be recovered from stabilizer measurements is very different. Hence arises the possibility of exploiting different graph states with equivalent computational power in order to better characterize the error sources.

\section{Promise settings} \label{sec:promise_settings}

The problem as specified so far is underdetermined, since we need to determine $3N$ parameters, i.e.\ the field directions and intensities, from just $N$ parameters, i.e.\ from the measured probability differences. Thus we need to modify the problem to make it possible to reconstruct the stray fields. We will therefore first consider simplified settings, in which the fields fulfill a condition (a \emph{promise}) that reduce the number of parameters to be estimated from $3N$ to $N$ -- thus permitting to estimate them through the measurement of the correlators stabilizing the graph state. We will then consider a different scenario in Sec.~\ref{sec:extension}, where we will extend the set of stabilizer measurements in order to collect information on all the $3N$ parameters.

From now on, we will assume that the field is always globally pointing in a known and well-determined direction, but has unknown site-dependent intensity. Specifically, we investigate the cases in which the field are parallel to one of the Cartesian axes; under these assumptions, Eq.~\eqref{general} simplifies considerably:
\begin{enumerate}
\item[1.] \textbf{Z-field} $[\ n_x = 0,\ n_z = 1,\ \beta_a =\, ?\ ]$\\ the field is always aligned in the $\hat{\boldsymbol{z}}$ direction
\begin{equation} \label{Z}
   \DP_a =  \beta_a \,;
\end{equation}
\item[2.] \textbf{X-field} $[\ n_x = 1,\ n_z = 0,\ \beta_a =\, ?\ ]$\\ the field is always aligned in the $\hat{\boldsymbol{x}}$ direction
\begin{equation} \label{X}
  \DP_a = \prod_{b \in \mathcal{N}_a} \beta_b \,;
\end{equation} 
\item[3.] \textbf{Y-field} $[\ n_x = 0,\ n_z = 0,\ \beta_a =\, ?\ ]$\\ the field is always aligned in the $\hat{\boldsymbol{y}}$ direction
\begin{equation} \label{Y}
  \DP_a = \beta_a \prod_{b \in \mathcal{N}_a} \beta_b \,.
\end{equation}
\end{enumerate}

\section{Solving methods} \label{sec:solving_methods}

Here we will present a method that allows us to infer the stray fields, knowing the probability differences $\DP_a $, in the promise settings specified above. We will show that not in all circumstances it is possible to determine all field intensities $\beta_a$, and this impossibility depends non-trivially on the connectivity structure of the graph state.

To determine the fields we need to solve a system of $N$ polynomial equations (one for each $\DP_a$) in $N$ variables (one for each field intensity $\beta_a$) of degree at most $d$, where $d$ is the maximum number of neighbours of any vertex in the graph. In general, a system of polynomial equations can be solved via computation of the Gr\"obner basis associated to the polynomials. There exist well-known algorithms that compute Gr\"obner bases~\cite{Groebner1965,Faugere1999}, but the time complexity in the worst case scenario grows doubly-exponentially in $N$~\cite{Bardet2005}. However, with the method that we will introduce the particular polynomial equations arising from Eqs.~\mbox{(\ref{Z}--\ref{Y})} can be solved efficiently. Notice in particular that the case of Eq.~\eqref{Z} is already trivially solved, as one reads out directly the (cosine of the) intensity of the field from the measured rates. The other two cases, Eqs.~\mbox{(\ref{X},~\ref{Y})}, will be discussed in the following sections.

\subsection{Complex logarithms} \label{sec:logarithms}

Consider the settings in which the stray field is parallel to either the $\hat{\boldsymbol{z}}$, $\hat{\boldsymbol{x}}$ or $\hat{\boldsymbol{y}}$ axis. We can write Eqs.~\mbox{(\ref{Z}--\ref{Y})} as 
\begin{equation} 
  \prod_{b \in \mathcal{N}_a^s} \beta_b = \DP_a \,,
\end{equation}
with $s \in \{x,y,z\}$, and setting $\mathcal{N}_a^z = \{a\}$ for the $Z$-field case, $\mathcal{N}_a^x = \mathcal{N}_a$ for the $X$-field case, and $\mathcal{N}_a^y = \mathcal{N}_a \cup \{a\}$ for the $Y$-field case. They are polynomial equations in the variables $\beta_a$, in which each polynomial has only two terms, i.e., the monomials appearing on the left hand and on the right hand side. To solve these we take the logarithms of both sides of the equations.

In general the cosines of the fields ($\beta_a$) and the probability differences ($\DP_a$) have support in the interval $[-1,1]$. Thus we have to use the logarithm as a complex-valued function; more precisely, we define the logarithm as taking values on the Riemann surface $\mathds{C}/_{2 \pi i}$:
\begin{equation}
\begin{array}{ll}
  \ln : \mathds{C}^\ast & \rightarrow \mathds{C}/_{2 \pi i} \\
  \ln : z               & \mapsto    \ln(|z|) + i \, \arg(z)
\end{array}
\end{equation}
where $\ln(|z|)$ is the real logarithm (mapping $\mathds{R}^+$ in $\mathds{R}$), $\mathds{C}^\ast = \mathds{C} \setminus \{0\}$ and the Riemann surface $\mathds{C}/_{2 \pi i}$ is the complex plane modulo identification of numbers that differ by an integer multiple of $2 \pi i$; we also denote by $\equal$ the equality on this Riemann surface. If we restrict $z$ to be a real number, then $(i \, \arg(z))$ is in the set $\{0, i \pi \}$. With this definition the logarithm is a single-valued function and the multiplication rule applies:
\begin{equation}
  \ln(z_1 z_2) \equal \ln(z_1) + \ln(z_2) \qquad \forall \ z_1,z_2 \in \mathds{C}^\ast \,.
\end{equation}
Notice that on the Riemann surface $\mathds{C}/_{2 \pi i}$ the division is not uniquely defined. In particular, when performing a division by $m \in \mathds{Z} \setminus \{0\}$ we get $|m|$ different valid results:
\begin{equation}\label{log_multiple_sol}
  z \div m \equal \left\lbrace \, \left. \frac{z}{m} + \frac{2k\pi}{m}i \  \right| \, k \in \{0,\ldots,|m|-1\} \, \right\rbrace  \,.
\end{equation}

\subsection{Linear equations} \label{sec:linear_eq}

Thus, after taking the logarithm of both sides of Eqs.~\mbox{(\ref{Z}--\ref{Y})}, we obtain the following linear equations:
\begin{equation} \label{linear_eq}
\underbrace{
\left(
  \begin{array}{c}
  \\
  \text{Adjacency}\\
  \text{Matrix} \\
  \\
  \end{array}
\right)
}_{\textstyle A_s}
\left(
  \begin{array}{c}
  \ln \beta_1 \\
  \ln \beta_2 \\
  \vdots      \\
  \ln \beta_N
  \end{array}
\right) \equal
\left(
  \begin{array}{c}
  \ln \DP_1 \\
  \ln \DP_2 \\
  \vdots         \\
  \ln \DP_N
  \end{array}
\right) \,.
\end{equation}
The adjacency matrix $A$ of a graph $G$ is a symmetric matrix defined as
\begin{equation}
[A]_{a, b} := 
\begin{cases}
  1 & \text{if} \ (a,b)    \in E \\
  0 & \text{if} \ (a,b) \notin E 
\end{cases} .
\end{equation}
We also define $A_s$ for $s \in \{x,y,z\}$ as 
\begin{align}
  A_z & := \one \white{+A} \qquad \text{$Z$-field case} \nonumber\,,\\
  A_x & := A \white{+\one} \qquad \text{$X$-field case} \,,\\
  A_y & := A + \one        \qquad \text{$Y$-field case} \nonumber\,, 
\end{align}
where $\one$ denotes the $N \times N$ identity matrix.

\subsection{Multiple solutions} \label{sec:multiple}

The problem thus has been reduced to solving a linear system of the form $A_s \boldsymbol{v} \equal \boldsymbol{w}$, with $\boldsymbol{v} := (\ln \beta_1, \ldots , \ln \beta_N)^T$ and $\boldsymbol{w} := (\ln \DP_1, \ldots , \ln \DP_N)^T$. From here, provided that the matrix $A_s$ is non-singular, the (logarithm of the cosines of the) fields can be reconstructed via linear algebra. However, in this context even when $A_s$ is non-singular it is not formally correct to write the solution as $\boldsymbol{w} = (A_s)^{-1} \boldsymbol{v}$, since the system $A_s \boldsymbol{v} \equal \boldsymbol{w}$ can have multiple solutions. This happens because the entries of $A_s$ are taken as elements of $\mathds{C}/_{2 \pi i}$, thus divisions by integer number have to be performed according to Eq.~\eqref{log_multiple_sol}. Therefore, in order recover all the valid solutions to the linear system~\eqref{linear_eq} we need to employ Eq.~\eqref{log_multiple_sol} whenever performing a division during the solving algorithm.

We adopt a modified version of Gauss elimination algorithm which allows us to recover these multiple solutions. Since the entries of $A_s$ are all integers, it is possible to bring $A_s$ into the form of an upper triangular matrix $Q$, applying only a sequence of the following \emph{elementary Gauss operations} on $A_s$:
\begin{description}
\item[Rule 1] multiply a row of $A_s$ by $-1$;
\item[Rule 2] permute two rows of $A_s$;
\item[Rule 3] add to the $l$-th row of $A_s$ an integer multiple of the $m$-th row, with $m\neq l$.
\end{description}
These operations can in fact be used to compute the \emph{Hermite normal form}~\cite{Hermite1851} 
of $A_s$. That is, it is always possible to write $A_s$ in the form:
\begin{equation}
  A_s = P \ Q \,,
\end{equation}
where $P$ and $Q$ are matrices with integer entries, $P$ unimodular ($|\det P| = 1$) and $Q$ upper-triangular; for more details see, e.g., Ref.~\cite{Kannan1979}. Notice that, $P$ being unimodular, $P^{-1}$ is also unimodular and has integer entries; moreover, $P^{-1}$ is obtained via multiplication of the matrices that represent the elementary Gauss operations. Hence the solutions of $A_s \boldsymbol{v} \equal \boldsymbol{w}$ are in one-to-one correspondence with the solutions of
\begin{equation}
  Q \, \boldsymbol{v} \equal P^{-1} \boldsymbol{w} \,.
\end{equation}

Thus Eqs.~\eqref{linear_eq} have been brought into the form $Q\,\boldsymbol{v} \equal \boldsymbol{w}'$, with $\boldsymbol{w}':= P^{-1} \boldsymbol{w}$, and $Q$ upper-triangular. If $A_s$ is non-singular, then all the diagonal entries $(Q_{11}, \ldots , Q_{NN})$ of $Q$ are different from $0$. Therefore the solutions can be recovered finding the $|Q_{NN}|$ solutions (see Eq.~\eqref{log_multiple_sol}) to the equation  $Q_{NN} v_N \equal w'_N$, then substituting the obtained values for $v_N$ in Eq.~\eqref{linear_eq}, then solving for $v_{N-1}$ and so forth. That is, having computed the possible values for $(v_{l+1}, \ldots, v_{N} )$ we obtain $v_l$ by solving the equation $Q_{ll} v_l \equal w'_l - \sum_{j=l+1}^N Q_{lj} v_j$. In total, the number of complex solutions is given by $\prod_{l=1}^N |Q_{ll}| = |\det\, A_s|$.

An equivalent way of looking at this operation, which will become useful in the error analysis in Sec.~\ref{sec:error_propagation}, is to consider the vector $\boldsymbol{w}'$ as taking $|\det\, A_s|$ different values, and then computing the corresponding solutions for $\boldsymbol{v}$ with standard linear algebra over the complex numbers. Explicitly, we need to solve $|\det\, A_s|$ different linear equations in the form $Q \boldsymbol{v} = \boldsymbol{w}_{c_1 \ldots c_N}'$ over $\mathds{C}$, in which the known terms are given by
\begin{equation} \label{indices}
  \boldsymbol{w}_{c_1 \ldots c_N}' := \boldsymbol{w}' + (2\pi i) \cdot \sum_{l=1}^N c_l \,\hat{\boldsymbol{e}}_l \,,
\end{equation}
where $\hat{\boldsymbol{e}}_l$ is the unit vector with a $1$ in position $l$, and $c_l \in \{0, \ldots, |Q_{ll}|-1\}$, $\forall l \in \{1, \ldots, N\}$.

In conclusion, we face two possible scenarios:
\begin{enumerate}
\item The matrix $A_s$ is singular; the system~\eqref{linear_eq} is either inconsistent or underdetermined.
\item The matrix $A_s$ is non-singular; the number of complex solutions is given by $|\det \, A_s|$.
\end{enumerate}

Finally, once the solutions $\boldsymbol{v}$ to the system~\eqref{linear_eq} have been computed, the corresponding values of the parameters $\beta_a$ can be obtained via element-wise exponentiation of $\boldsymbol{v}$.

\subsection{Constraints on the solutions} \label{sec:constraints}

Now consider $\boldsymbol{w} = (\ln \DP_1, \ldots , \ln \DP_N)^T$; the components of $\boldsymbol{w}$ are logarithms of real numbers, thus the imaginary part must be an integer multiple of $\pi$ (i.e.\ the imaginary component is either $0$ or $i \pi$ on $\mathds{C}/_{2\pi i}$). After the Gauss elimination algorithm, we obtain another vector $\boldsymbol{w}' = P^{-1} \boldsymbol{w}$ whose entries are integer linear combinations of those of $\boldsymbol{w}$; thus, the entries of $\boldsymbol{w}'$ still have an imaginary part which is an integer multiple of $\pi$. 

Take $v_l$ as a complex solution of an equation of the form $Q_{ll} v_l \equal w'_l - \sum_{j=l+1}^N Q_{lj} v_j$; if $Q_{ll}$ is odd, then among these solutions only one has an imaginary part which is an integer multiple of $\pi$; if $Q_{ll}$ is even, then among these solutions only two have imaginary parts which are integer multiples of $\pi$. Repeating this argument for all entries of $\boldsymbol{v}$ we get the number of allowed solutions is $2^\mu$, where $\mu$ is the number of even diagonal entries in $Q$. 

The requirement that all the probabilities and the (cosines of the) field strengths must be real thus restricts the allowed solutions to just a small subset of all possible complex solutions. There is still another requirement that should be enforced: the parameters $\beta_a$ and $\DP_a$ have to lie in the range $[-1,+1]$; this condition further restricts the number of legitimate solutions. Notice that, when we enforce this requirement, the number of acceptable solutions cannot be expressed in terms of properties of the adjacency matrix alone ($A_s$), since this number does depend on the data actually collected ($\boldsymbol{w}$).

\section{Solution for particular classes of graph states} \label{sec:particular_classes}

In this section, we will explicitly determine whether a graph state allows a complete reconstruction of the fields acting at each site -- assuming that it is parallel to either the $\hat{\boldsymbol{x}}$, $\hat{\boldsymbol{y}}$ or $\hat{\boldsymbol{z}}$ direction -- for some relevant classes of graphs. We will consider at the beginning graphs which represent one-dimensional lattices -- that is, \emph{linear chains}. In these graphs each vertex is connected to at most two neighbours. Afterwards, we will show how to extend those results to two-dimensional (and higher dimensional) square lattices. Finally, we will give an application for states used in quantum repeaters and measurement-based approaches to quantum error correction.

\subsection{Linear chains} \label{sec:linear_chains}

A graph represents a \emph{closed linear chain} if all vertices are connected to other two vertices in the graph; it represents a \emph{open-ended linear chain} if there are two endpoints which are connected to only one other vertex. 

The adjacency matrix of an open-ended linear chain with $N=m$ vertices is a $m \times m$ matrix given by:
\begin{equation}\label{A_open}
[A^{open}]_{a,b} =
  \begin{cases} 1 & \text{if } |a-b|= 1 \\
   0              & \text{otherwise}
  \end{cases}
  \qquad \forall \, a,b \in [m]
\end{equation} 
in which we introduce the notation $[m] := \{1,\ldots, m\}$ and we have numbered the vertices from $1$ to $m$ in the natural way (from one endpoint to the other). Thus $A_x^{open} \equiv A^{open}$ and $A_y^{open} \equiv A^{open}+\one$ are both tridiagonal Toeplitz matrices -- and where a Toeplitz matrix $T$ is a constant-diagonal matrix, i.e.\ $[T]_{a,b} = t_{(a-b)}$, $\forall \ a,b \in [m]$. Their determinants can be given as closed-form expressions~\cite{Noschese2013}:
\begin{align}
\det (A_x^{open}) & = 
  \begin{cases}
  0  & \text{if } m \equiv 1 \mod 2 \\
  +1 & \text{if } m \equiv 0 \mod 4 \\
  -1 & \text{if } m \equiv 2 \mod 4 
  \end{cases} \\
\det (A_y^{open}) & =
  \begin{cases}
  0  & \text{if } m \in  \{2,5\} \mod 6 \\
  +1 & \text{if } m \in  \{0,1\} \mod 6 \\
  -1 & \text{if } m \in  \{3,4\} \mod 6 \,.
  \end{cases} 
\end{align}
Therefore, given the solution methods presented in Sec.~\ref{sec:solving_methods}, it is evident that the system is underdetermined when $m$ is even in the case of $X$-fields, and when $m \equiv 2 \ {(\text{mod } 3)}$ for $Y$-fields. In all other cases there is exactly one (complex) solution to the equations.

For a closed linear chain with $m$ vertices the adjacency matrix is given by:
\begin{align} \label{A_close}
[A^{close}]_{a,b} =
  & \begin{cases} 
   1 & \text{if } |a-b| \in \{1,m-1\} \\
   0 & \text{otherwise}
  \end{cases} \\
  & \qquad \ \forall \, a,b \in [m] \,, \nonumber
\end{align}
and thus $A_x^{close}$ and $A_y^{close}$ are circulant matrices -- where a circulant matrix $C$ is defined by the relation $[C]_{a,b} = c_{(a-b) \text{ mod } m}$, $\forall \ a,b \in [m]$. Also in this case it is possible to give closed-form expressions for their determinants~\cite{Gray1971}, but we have to take into account that $m=1$ and $m=2$ are special cases:
\begin{align}
\det (A_x^{close}) & = 
  \begin{cases}
    \begin{cases}
     0 & \text{if } m=1 \\
    -1 & \text{if } m=2
    \end{cases} \\
    \begin{cases}
       & \text{If } m \geq 3 \text{ and:} \\
    +2 & \text{if } m \equiv 1 \mod 2 \\
     0 & \text{if } m \equiv 0 \mod 4 \\
    -4 & \text{if } m \equiv 2 \mod 4
    \end{cases}
  \end{cases} \\
\det (A_y^{close}) & = 
  \begin{cases}
    \begin{cases}
    +1 & \text{if } m=1 \\
     0 & \text{if } m=2
    \end{cases} \\
    \begin{cases}
       & \text{If } m \geq 3 \text{ and:} \\
     0 & \text{if } m \in \{0,3\}  \mod 6 \\
    +3 & \text{if } m \in \{2,4\}  \mod 6 \\
    -3 & \text{if } m \in \{1,5\}  \mod 6  \,.
    \end{cases}
  \end{cases}
\end{align}
Therefore, the problem is underdetermined when $m=1$ or $m \equiv 0 \ {(\text{mod } 4)}$ in the case of $X$-fields, and when $m=2$ or $m \equiv 0 \ {(\text{mod } 3)}$ in the case of $Y$-fields. In the other cases, there are a number of complex solutions equal to $|\det (A_s^{close})|$. For the $Y$-field case, since the non-zero determinant of $A_y^{close}$ is equal to $+3$ or $-3$, there is only one real solution for the fields. For the $X$-field case, when $m \equiv 1 \ {(\text{mod } 2)}$, $\det (A_x^{close}) = 2$, so there are two real solutions; and when $m \equiv 1 \ {(\text{mod } 2)}$, $\det (A_x^{close}) = 4$, and it can be verified that the Hermite normal form $Q$ has two diagonal entries equal to $2$, so there are four admissible solutions. A summary of these results can be found in Table~\ref{table1}.

For reference, we also give the eigenvalues of the adjacency matrices $A_s^{open}$ and $A_s^{close}$ for $s \in \{x,y\}$:
\begin{align}
  \lambda_j^{(x, open)} & = 2 \cos \left( \frac{\pi j}{m+1} \right) \qquad \forall \, j \in \{1, \ldots , m \}\\
  \lambda_j^{(x, close)}& = 2\,\widetilde{\cos}_m \left( \frac{2 \pi j}{m} \right) \  \qquad \forall \, j \in \{1, \ldots , m \}\\
  \lambda_j^{(y, open)} & = \lambda_j^{(x, open)} + 1 \qquad \quad \ \forall \ j,m \\
  \lambda_j^{(y, close)}& = \lambda_j^{(x, close)}+ 1 \qquad \quad \ \forall \ j,m
\end{align}
in which we have introduced the function $\widetilde{\cos}_m(\theta)$ to deal with the special cases arising when $m=1$ or $m=2$:
\begin{equation}
  \widetilde{\cos}_m(\theta) :=
    \begin{cases}
     0              & \text{if } m=1 \\
    -\cos(\theta)/2 & \text{if } m=2 \\
    \cos(\theta)    & \text{if } m\geq 3 \ .
    \end{cases} 
\end{equation}
The results above will turn out useful in the next section.

\begin{table}
  \begin{ruledtabular}
  \begin{tabular}{ll}  
  \textbf{Open-end linear chain} & 
  \textbf{Closed linear chain}\\
  \hspace*{3mm}\includegraphics[scale=.20]{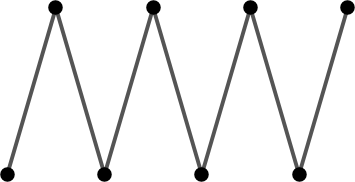} &
  \hspace*{6mm}\includegraphics[scale=.20]{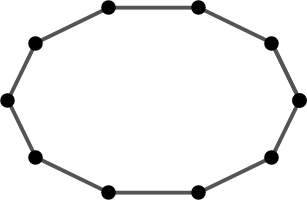} \\
  \hline 
  \textbf{Adjacency matrix:} & 
  \textbf{Adjacency matrix:}\\
  tridiagonal Toeplitz matrix &
  tribanded circulant matrix
  \vspace*{2mm}\\   
    $\left(
     \begin{array}{ccccc}
      0 & 1 & 0 &\cdot& 0\\
      1 & 0 & 1 &\cdot& 0\\
      0 & 1 & 0 &\cdot&\cdot \vspace*{-1mm} \\
    \cdot&\cdot&\cdot&\ddots&1\\
      0 & 0 &\cdot&1& 0 
      \end{array}
      \right)$  &
    $\left(
     \begin{array}{ccccc}
      0 & 1 & 0 &\cdot& 1\\
      1 & 0 & 1 &\cdot& 0\\
      0 & 1 & 0 &\cdot&\cdot \vspace*{-1mm}\\
    \cdot&\cdot&\cdot&\ddots&1\\
      1 & 0 &\cdot&1& 0 
     \end{array}
     \right)$
     \vspace*{2mm}\\
  \textbf{Eigenvalues:} & 
  \textbf{Eigenvalues:}\\
  $ \lambda_j^{(x)} = 2 \cos \left( \frac{\pi j}{m+1} \right)$ &
  $ \lambda_j^{(x)} = 2\,\widetilde{\cos}_m \left( \frac{2 \pi j}{m} \right)$ \\
  $ \lambda_j^{(y)} = \lambda_j^{(x)} + 1$ &
  $ \lambda_j^{(y)} = \lambda_j^{(x)}+ 1$ 
  \vspace*{2mm}\\
  \multicolumn{2}{l}{\textbf{The adjacency matrix is not invertible when:}} \\
  \textbf{X-field:} &
  \textbf{X-field:} \\
   $\quad m \equiv 1 \mod 2$ &
   $\quad
    \begin{array}{l}
    m=1 \,\ (\text{for } m<3)\\
    m \equiv 0 \mod 4
    \end{array}$\\
  \textbf{Y-field:} &
  \textbf{Y-field:} \\
   $\quad m \equiv 2 \mod 3 $ &
   $\quad
    \begin{array}{l}
    m=2 \,\ (\text{for } m<3)\\
    m \equiv 0 \mod 3
    \end{array}$ 
  \end{tabular}
  \end{ruledtabular}
  \caption{Summary of the features of the adjacency matrices of graphs representing linear chains with open and closed boundary conditions.}
  \label{table1}
\end{table}

\subsection{Square lattices}

Here we describe the method for solving the problem for two dimensional square lattices. The same constructions can be applied straightforwardly in an arbitrary number of dimensions, so in the end we will present the formulas also for the general multidimensional case.

As a first step, we give the expression for the adjacency matrix of a graph representing a two-dimensional square lattice, having $m_1$ rows in one direction and $m_2$ columns in the other direction, thus having in total $N = m_1 \cdot m_2$ vertices. We number the vertices according to a scan-line pattern, i.e.\ we assign to the vertex in position $(j_1,j_2)$, with $j_1 \in [m_1]$ a row index and $j_2 \in [m_2]$ a column index, the value $j_2 + (j_1-1)m_2$. With this numbering convention, the adjacency matrices $A$ of planar lattices take the form of block-Toeplitz matrices. That is, there are $m_1 \times m_1$ blocks, each one consisting of $m_2 \times m_2$ elements, and the blocks which lie on the same diagonal are all equal; the blocks themselves are Toeplitz matrices.

For a square lattice with open boundaries in both directions, i.e.\ for a \emph{planar graph}, the adjacency matrix takes the following tridiagonal block-matrix form:
\begin{align} \label{A_planar}
[A^{planar}]_{a,b} = &
    \begin{cases}
    A^{open}_{m_2}  & \text{if } |a-b| = 0 \\
    \one_{m_2}      & \text{if } |a-b| = 1 \\
    0_{m_2}         & \text{otherwise}
    \end{cases} \\
& \qquad \qquad \quad \forall \ a,b \in [m_1] \nonumber
\end{align}
where here $a,b$ are block indices, the blocks are of size $m_2 \times m_2$, and $A^{open}_{m_2}$ is given by Eq.~\eqref{A_open}. 

Now we consider another family of graphs, which are obtained from the above planar square lattice by connecting the vertices of the $m_2$-th column to the corresponding vertices in the first column; so we obtain a graph which is topologically a cylinder. Then the adjacency matrix takes the following block-matrix form:
\begin{align} \label{A_cylinder}
[A^{cylinder}]_{a,b} = &
    \begin{cases}
    A^{close}_{m_2} & \text{if } |a-b| = 0 \\
    \one_{m_2}      & \text{if } |a-b| = 1 \\
    0_{m_2}         & \text{otherwise}
    \end{cases} \\
& \qquad \qquad \quad \forall \ a,b \in [m_1] \nonumber
\end{align}
where $A^{close}_{m_2}$ is given by Eq.~\eqref{A_close}.

Finally, we consider the family of graphs obtained from the cylindrical graph by connecting the vertices of the $m_1$-th row to the corresponding vertices of the first row; so we obtain a graph which is topologically a torus. The adjacency matrix takes the following block-matrix form:
\begin{align} \label{A_torus}
[A^{torus}]_{a,b} = &
    \begin{cases}
    A^{close}_{m_2} & \text{if } |a-b| = 0 \\
    \one_{m_2}      & \text{if } |a-b| \in \{1,m_1 - 1\} \\
    0_{m_2}         & \text{otherwise}
    \end{cases} \\
& \qquad \qquad \quad \forall \ a,b \in [m_1] \nonumber
\end{align}
with the same conventions as above. 

As usual, we have $A_x = A$ and $A_y = A + \one$ for all the cases above, where $\one$ is the $m_1m_2 \times m_1m_2$ identity matrix.

Now we will show how to obtain the eigenvalues of the above matrices in closed form. Once the eigenvalues are given, it is straightforward to tell whether all information about the fields can be recovered or not, by searching for eigenvalues equal to zero. If these can be found, it means that the adjacency matrix is singular and not all the information can be recovered. The number of zero eigenvalues is the \emph{rank defect} of the adjacency matrix, and is a measure of the amount of information that one cannot reconstruct.

\begin{table}
  \begin{ruledtabular}
  \begin{tabular}{ll}  
  \textbf{Tridiagonal}  & 
  \textbf{Tribanded} \\
  \textbf{Toeplitz matrix} &
  \textbf{circulant matrix}
  \vspace*{2mm}\\   
    $\left(
     \begin{array}{ccccc}
      s & t & 0 &\cdot& 0\\
      u & s & t &\cdot& 0\\
      0 & u & s &\cdot&\cdot \vspace*{-1mm} \\
    \cdot&\cdot&\cdot&\ddots&t\\
      0 & 0 &\cdot & u & s 
      \end{array}
      \right)$  &
    $\left(
     \begin{array}{ccccc}
      s & t & 0 &\cdot& t\\
      t & s & t &\cdot& 0\\
      0 & t & s &\cdot&\cdot \vspace*{-1mm} \\
    \cdot&\cdot&\cdot&\ddots&t\\
      t & 0 &\cdot & t & s 
      \end{array}
    \right)$
     \vspace*{2mm}\\
  \textbf{Eigenvalues:} & 
  \textbf{Eigenvalues:}\\
  $ \lambda_k^{(T)} = s + 2\sqrt{tu} \ \cos \left( \frac{\pi k}{m+1} \right) $ &
  $ \lambda_k^{(C)} = s + 2 t \ \widetilde{\cos}_m \, \left(\frac{2\pi k}{m} \right) $ 
  \end{tabular}
  \end{ruledtabular}
  \caption{Eigenvalues of tridiagonal Toeplitz matrices and of tribanded symmetric circulant matrices.}
  \label{table2}
\end{table}

To solve our problem, we study tridiagonal Toeplitz matrices and circulant matrices with three bands, of size $m \times m$. A tridiagonal Toeplitz matrix $T$ has the form:
\begin{equation} \label{Toeplitz}
[T]_{a,b} = 
    \begin{cases}
    s  & \text{ if } a-b = 0 \\
    t  & \text{ if } a-b = +1 \\
    u  & \text{ if } a-b = -1 \\
    0  & \text{ otherwise}
    \end{cases}
    \qquad \forall \, a,b \in [m]
\end{equation}
and its eigenvalues are given by the expression~\cite{Noschese2013}:
\begin{equation} \label{eigen_Toeplitz}
  \lambda_{k}^{(T)} = s + 2\sqrt{tu} \ \cos \left( \frac{\pi k}{m+1} \right) \quad \forall \, k \in [m] \,.
\end{equation}
By standard spectral methods, it is straightforward to show that this expression is meaningful also when $s, t, u$ are pairwise commuting diagonalisable matrices. Under these assumption, the matrices are simultaneously diagonalisable, and thus it is sufficient to define the necessary operations as acting over the eigenvalues of these matrices as given in the common eigenvector basis. 

Similarly, a circulant matrix $C$ has the form:
\begin{equation} \label{circulant}
  [C]_{a,b} = c_{(a-b) \text{ mod } m} \qquad \forall \, a,b \in [m] \,,
\end{equation}
for an arbitrary set of coefficients $(c_0, \ldots, c_{m-1})$; to specialise to the case in which there are only three bands, it is sufficient to set $c_j \equiv 0$ for $j \notin \{-1,0,+1\}$. The eigenvalues of $C$ are then given by~\cite{Gray1971}:
\begin{equation} \label{eigen_circulant}
  \lambda_{k}^{(C)} = \sum_{j=0}^{m-1} \, c_j \ \omega_{k}^{\ j} \qquad \forall \, k \in [m] \,,
\end{equation}
where $\omega_{k} := \exp (2\pi i k / m)$ are the roots of unity. The result can be specialised to the case in which $s \equiv c_0$, $t\equiv c_{-1} = c_1$, with $t = t^\dag = t^\ast$ and $c_j = 0$ for $j \notin \{-1,0,+1\}$, giving the formula:
\begin{equation}
  \lambda_{k}^{(C)} = s + 2 t \, \widetilde{\cos}_m \, \left(\frac{2\pi k}{m} \right) \quad \forall \, k \in [m] \,.
\end{equation}
As we argued before, again this formula is valid also when $s, t$ are commuting diagonalisable matrices. A summary of these properties can be found in Table~\ref{table2}.

\begin{figure*}
  \begin{ruledtabular}
  \begin{tabular}{ccc}
    \includegraphics[scale=0.24]{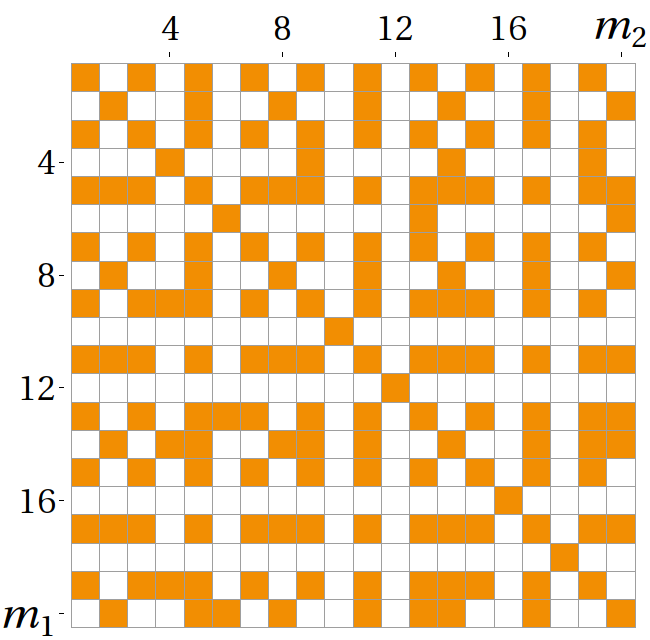} &
    \includegraphics[scale=0.24]{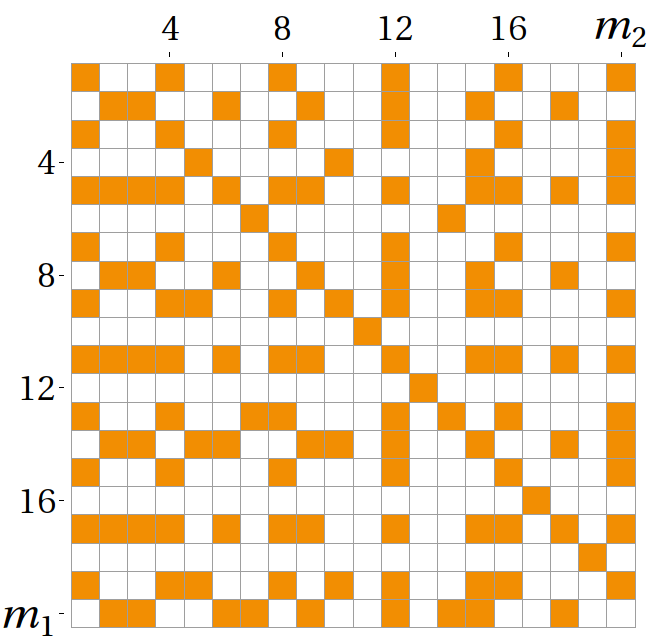} &
    \includegraphics[scale=0.24]{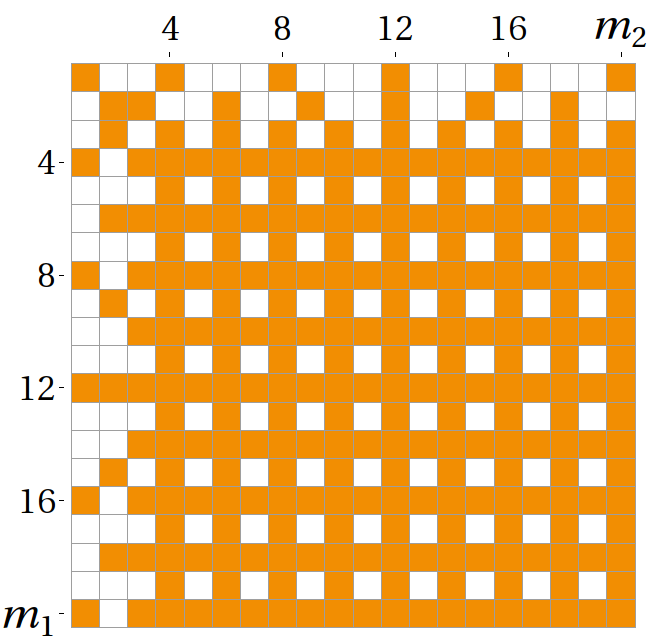} \\
    $X$-field, planar &
    $X$-field, cylindrical &
    $X$-field, toroidal
    \vspace*{3mm}\\    
    \includegraphics[scale=0.24]{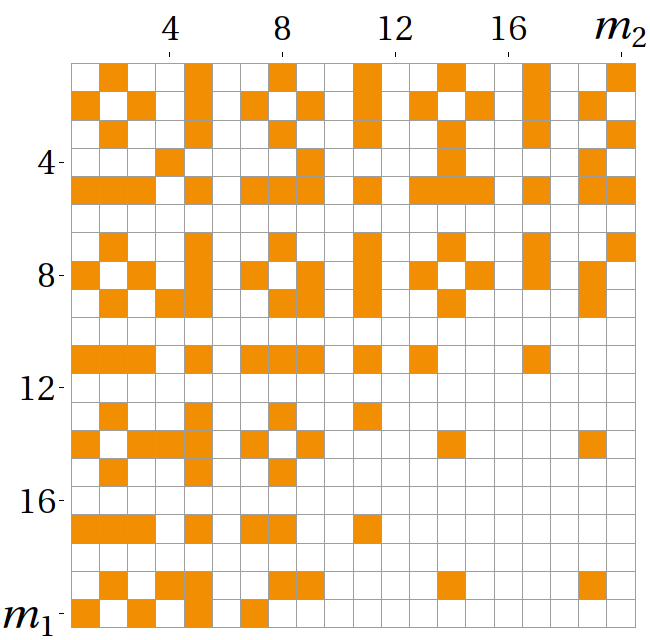} &
    \includegraphics[scale=0.24]{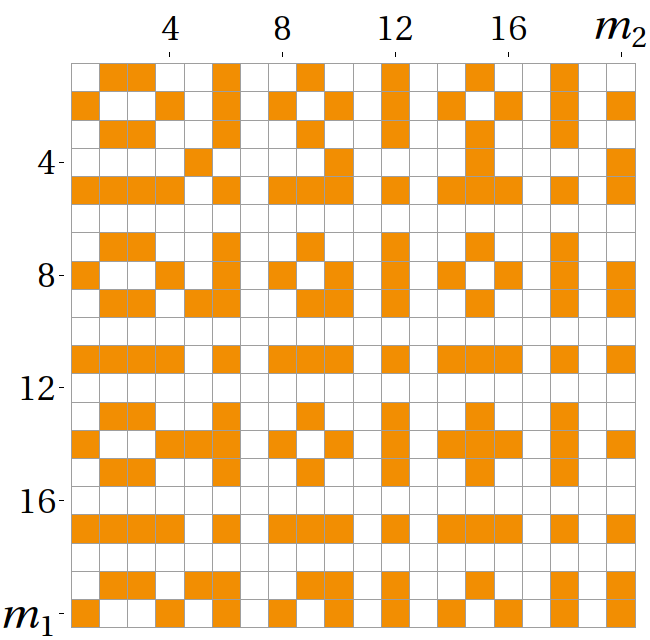} &
    \includegraphics[scale=0.24]{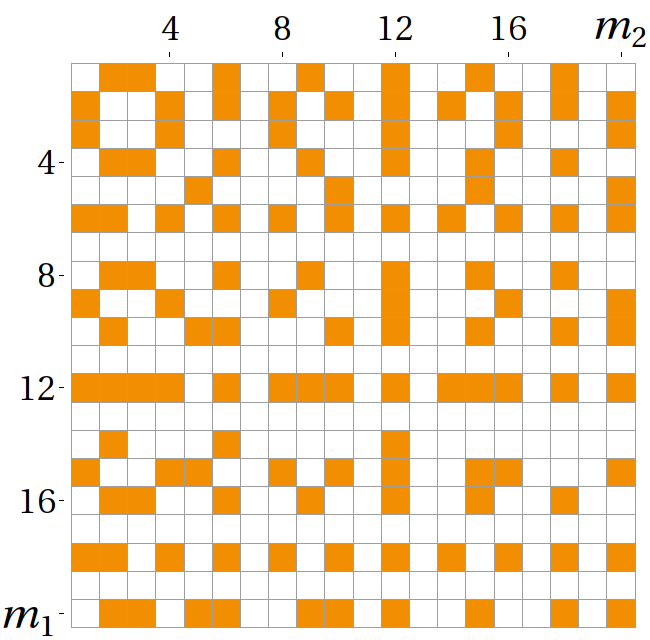} \\
    $Y$-field, planar &
    $Y$-field, cylindrical &
    $Y$-field, toroidal\\
  \end{tabular}
  \end{ruledtabular}
  \caption{Summary of fully reconstructible fields. Each cell in the arrays corresponds to a different square lattice. In each array the entry in position $(m_1,m_2)$ is filled if the adjacency matrix of a square-lattice graph having $m_1 \times m_2$ vertices is singular. In the case of cylindrical geometry, the open boundary condition is in the direction having $m_1$ rows and the closed boundary condition is in the direction having $m_2$ columns. }
  \label{fig0}
\end{figure*}

At this point we are ready to compute all the eigenvalues of the adjacency matrices $A^{planar}$, $A^{cylinder}$ and $A^{torus}$. The matrix $A^{planar}$, as given by Eq.~\eqref{A_planar}, is a block tridiagonal Toeplitz matrix, thus its (block) eigenvalues are given by Eq.~\eqref{eigen_Toeplitz}, with the identifications $s = A_{m_2}^{open}$ and $t=u= \one_{m_2}$:
\begin{align} 
  \Lambda_{k_1}^{planar} = \, A_{m_2}^{open} \, + \, \one_{m_2} \times 2 \cos & \left( \frac{\pi k_1}{m_1 +1} \right)  \\
    & \ \forall \ k_1 \in [m_1] \,. \nonumber
\end{align}
These ``eigenvalues'' $\Lambda_{k_1}^{planar}$ are by themselves $m_2\times m_2$ tridiagonal Toeplitz matrices, whose eigenvalues can be computed using again formula~\eqref{eigen_Toeplitz}, this time with the identification $s = 2 \cos (\pi k_1 / (m_1 +1))$ and $t=u=1$. Finally, a (real) eigenvalue of $\Lambda_{k_1}^{planar}$ is also a (real) eigenvalue of $A^{planar}$, this time considering it as a matrix of real (or complex) numbers of size $m_1m_2 \times m_1m_2$. So, in conclusion, all the $m_1m_2$ eigenvalues of $A_x^{planar}$ are given by the expression:
\begin{align}
    \lambda_{k_1,k_2}^{planar} = \ & 2 \cos \left(\frac{\pi k_1}{m_1+1} \right) + 2 \cos\left( \frac{\pi k_2}{m_2+1} \right) \\
    & \forall \ k_1 \in [m_1], \ k_2 \in [m_2] \,. \nonumber
\end{align}
An analogous computation can be performed for $A^{cylinder}$ and $A^{torus}$, resulting in:
\begin{align}
    \lambda_{k_1,k_2}^{cylinder} & = \ 
            2 \, \widetilde{\cos}_{m_1}\left(\frac{2 \pi k_1}{m_1} \right)
         +  2\cos   \left(\frac{\pi k_2}{m_2 +1} \right) \\
\lambda_{k_1,k_2}^{torus} & = \
                       2 \, \widetilde{\cos}_{m_1}\left(\frac{2 \pi k_1}{m_1} \right) 
                     + 2 \, \widetilde{\cos}_{m_2}\left(\frac{2 \pi k_2}{m_2} \right)\\
    & \qquad \forall \ k_1 \in [m_1], \ k_2 \in [m_2] \,. \nonumber
\end{align}
Naturally, the eigenvalues of $A_y$ are equal to the eigenvalues $A_x$ plus one, for any adjacency matrix $A$. A list of the square lattice graph states of size up to $20 \times 20$ for which the error channel can be reconstructed is given in Fig.~\ref{fig0}.

\subsection{Cubic lattices in arbitrary dimension}

We conclude this section remarking that the procedure we have outlined carries through any number of dimensions. Therefore we can easily generalize the above results to a general cubic lattice in $d = d_1 + d_2$ dimensions, assuming closed boundary conditions in $d_1$ dimensions and open boundary conditions in $d_2$ dimensions. The eigenvalues of the adjacency matrix of these classes of cubic lattices are given by:
\begin{align} \label{d-dimensional}
  \lambda_{k_1, \ldots, k_d}^{(d_1,d_2)} & = 
    \sum_{r=1}^{d_1} 2\,\widetilde{\cos}_{m_r}\!\left(\frac{2 \pi k_r}{m_r} \right) 
  + \sum_{r=d_1+1}^{d_1+d_2} 2\cos \left(\frac{\pi k_r}{m_r+1} \right) \nonumber \\
        & \quad \forall \, k_r \in [m_r] \,, \ \forall \, r \in [d_1+d_2] \,,
\end{align}
where $m_r$ is the size of the lattice along dimension $r$.

\subsection{Application: quantum repeaters} \label{sec:repeaters}

In this section we provide an example of the application of the methods we have developed. We show that in many cases these techniques can be applied to states that are useful resources for measurement-based \emph{quantum repeaters} and measurement-based QEC.

A quantum repeater is a protocol that allows one to efficiently transmit quantum information over long distances in presence of noise that grows multiplicatively with the distance -- which happens, e.g.\ when transmitting photons along an optical fiber. To achieve this, it is necessary to add several intermediate stations along the transmission line; in these stations entanglement purification and entanglement swapping are applied to an ensemble of noisy Bell pairs in order to establish a single high-fidelity Bell pair between the distant parties~\cite{Wolfgang1999,Michael2012}. 

In the measurement-based approach to QEC a \emph{logical} qubit is encoded using a small size graph state as a resource. Then the information encoded in the logical qubit can be teleported to a freshly prepared low-error qubit entangling the graph state with this ancilla qubit and then performing local measurement on all the qubits of the original graph state. Alternatively, both entanglement purification and entanglement swapping can be performed at the same time exploiting again a local measurement pattern on specific graph states~\cite{Michael2014}.

In Refs.~\cite{Michael2012, Michael2015} some examples of graph states are given, up to local unitary ($LU$) corrections, that can be used for measurement-based quantum repeaters. Here we will not consider the $LU$-corrections, since these can always be included in the final read-out measurements; namely, we can effectively perform a $LU$-correction $U_a$ by measuring the (local) observable $U_a^\dag O_a U_a$ in place of the \emph{uncorrected} observable $O_a$. 

These graph-state resources include open-ended and closed linear chains of length $3$ and $5$; GHZ states with $4$ and $6$ qubits; and other, more irregular graph states. Since the linear chain cases have been already analysed in Sec.~\ref{sec:linear_chains}, we focus on the other cases.

We recall that a GHZ state with $N$ qubits can be represented as a graph state, up to $LU$-corrections, in at least two ways:
\begin{enumerate}
\item by a \emph{fully-connected} graph;
\item by a \emph{star} graph, i.e., a single qubit connected to all the other $N -1$ qubits. 
\end{enumerate} 
In the first representation the adjacency matrix $A^{(GHZ 1)}$ of a GHZ state has $0$ on the diagonal and $1$ in all other entries. We have that $\det \, (A^{(GHZ 1)}) = (-1)^{N-1}{(N-1)}$ for all $N \in \mathds{N}^\ast \equiv \mathds{N} \setminus \{0\} $ and that $\det \, (A^{(GHZ 1)} + \one) = 0$ for $N \neq 1$, thus it is possible to reconstruct the stray fields in the $X$-field case but not in the $Y$-field case. On the other hand, in the second representation the adjacency matrix $A^{(GHZ 2)}$ is a matrix with $1$ in position $(1,k)$ and $(k,1)$, for $k \in \{2,\ldots N\}$, and $0$ in all other entries. It can be shown that $\det \, (A^{(GHZ 2)}) = 0$ for $N\neq2$ and $\det \, (A^{(GHZ 2)} + \one) = 2-N$ for all $N$. In this case it is possible to reconstruct $Y$-fields but not $X$-fields. Therefore, by switching between the two representations, it is always possible to reconstruct the information for GHZ states, excluding only the trivial cases $N=1$ and $N=2$. These results are summarized in Table~\ref{table3}.

Another application of graph states is to implement measurement-based versions of QEC. For example the Steane $5$-qubit QEC code can be implemented using a graph state representing a closed linear chain with $5$ vertices together with a sixth vertex fully connected to all the other vertices. It can be checked that for this graph both the $X$-field and $Y$-field cases can be solved. Similarly, more complex graphs can be analysed case by case.

Finally, we remark that there is a very simple test to show in some cases that the $X$-field setting cannot be reconstructed. This happens whenever any two vertices in the graph have the same set of neighbours. In this circumstance two rows in the adjacency matrix are equal, and thus the matrix itself is singular.

\begin{table}
  \begin{ruledtabular}
  \vspace*{2mm}
  \begin{tabular}{cll}
  \raisebox{-.5\height}{\includegraphics[scale=0.11]{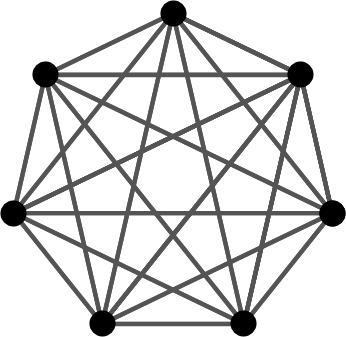}}
  &
  \raisebox{-.3\height}{
  $\ \begin{array}{l}
    \det (A_x^{GHZ 1}) = (-1)^{N-1} (N-1) \\
    \det (A_y^{GHZ 1}) = 0                
  \end{array}$ }
  &
  \raisebox{-.3\height}{
  $\begin{array}{l}
    \forall N \in \mathds{N}^\ast \\
    (N \neq 1) 
  \end{array}$  }
  \vspace*{2mm} 
  \\
  \raisebox{-.5\height}{\includegraphics[scale=0.10]{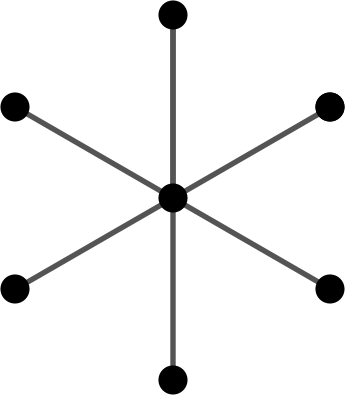}}
  &
  \raisebox{-.3\height}{
  $\ \begin{array}{l}
    \det (A_x^{GHZ 2}) = 0  \\
    \det (A_y^{GHZ 2}) = 2-N 
  \end{array}$ }
  &
  \raisebox{-.3\height}{
  $\begin{array}{l} 
    (N \neq 2)              \\
    \forall N \in \mathds{N}^\ast
  \end{array}$ }
  \end{tabular}
  \vspace*{2mm}
  \end{ruledtabular}
  \caption{Determinants of the adjacency matrices of the \emph{fully-connected} graph and of the \emph{star} graph.}
  \label{table3}
\end{table}

\section{Error propagation} \label{sec:error_propagation}

In this section we address the problem of how uncertainties in the probability differences $\DP_a$ and in the field alignments propagate through the reconstruction method we have given.

\subsection{Effect of uncertainty in the measured probabilities} \label{sec:effect_of_uncertainties}

First we consider the effect of the uncertainty arising from the fact that the correlator measurements are performed only a finite amount of times; thus the probability differences $\DP_a$ are known only up to a finite precision. For the moment however, we still assume that the stray fields are perfectly aligned to one of the axes ($\hat{\boldsymbol{x}}$, $\hat{\boldsymbol{y}}$ or $\hat{\boldsymbol{z}}$).  

Thus, each correlator $K_a$ at each vertex $a$ is measured $M$ times, yielding a string of results $(\kappa_a^{(1)}, \ldots, \kappa_a^{(M)}) \in \{0,1\}^M$. From these measurements, we can extract an \emph{error rate difference} $\Delta R_a \in [-1,+1]$, defined as the difference between the frequency of $\kappa_a=0$ outcomes and $\kappa_a=1$ outcomes:
\begin{equation} \label{rates}
  \Delta R_a := \frac{\#(\kappa_a = 0) - \#(\kappa_a = 1)}{M} \,.
\end{equation}
The random variables $\Delta R_a$ are sampled according to a shifted and rescaled version of the binomial distribution, and therefore they have mean and variance equal to:
\begin{equation}
  \E \big[ \Delta R_a \big] \,=\, \DP_a \,,
  \quad
  \E \big[ \big(\Delta R_a - \DP_a \big)^2 \big] \,=\, \frac{1-(\DP_a)^2}{M} \,.
\end{equation}
That is, $\Delta R_a$ are unbiased estimators of $\DP_a$. In the following, we will take the limit $M \gg 1$ and thus assume that $R_a$ is approximately distributed according to a normal probability distribution function with very small variance (a delta-like probability distribution function).

We then move to using estimators in logarithmic variables; i.e., we consider the estimator of $\ln (\DP_a)$ as given by $\omega_a \equiv w_a + \delta w_a := \ln (\Delta R_a)$, where $\boldsymbol{\omega} \rightarrow \boldsymbol{w}$ when $M\rightarrow \infty$, with the definition of $\boldsymbol{w}$ given in Sec.~\ref{sec:multiple}. The logarithm again takes values in $\mathds{C}/_{2\pi i}$. When the distribution of $R_a$ is delta-like and $\E (\Delta R_a) \neq 0$ (or, more correctly, $\E(\Delta R_a) \gg \text{Var} (\Delta R_a)$) it is legitimate to approximate the expectation value and variance of $w_a = \ln(R_a)$ with the following formulas:
\begin{align} 
  \E[\omega_a]         &\ \approx \ \ln(\E[\Delta R_a]) \\
  \label{variance}
  \text{Var}[\omega_a] &\ \approx \ \frac{\text{Var}[\Delta R_a]}{|\,\E[\Delta R_a]\,|} \,.
\end{align}

Then, through the reconstruction procedure given in Sec.~\ref{sec:solving_methods}, it is possible to give estimates of $\ln (\beta_a)$ in terms of estimates of $\ln (\DP_b)$. Namely, remembering that there are $|\det \, A_s|$ solutions to an equation in the form $A_s \boldsymbol{v} \equal \boldsymbol{\omega}$ (when $\det(A_s) \neq 0$), we set as estimator of $\ln (\beta_a)$ the value $v_a^{\,c} := \sum_b [A_s^{-1}]_{a,b} \, \omega_b^{\,c}$; here $s \in \{\hat{\boldsymbol{x}},\hat{\boldsymbol{y}},\hat{\boldsymbol{z}}\}$, $c \in \{1,\ldots, |\det\,A_s|\}$ and $A_s^{-1} \boldsymbol{\omega}^{\,c}$ denotes the $c$-th complex solution of the equation $A_s \boldsymbol{v} \equal \boldsymbol{\omega}$. We recall that this can be achieved by inverting the matrix $A_s$ in $\mathds{C}$ and then letting $\boldsymbol{\omega}$ assume $|\det \, A_s|$ different values $\boldsymbol{\omega}^{\,c}$, as shown in Eq.~\eqref{indices} (the index $c$ is in a one-to-one correspondence with the indices $(c_1,\ldots,c_N)$), which then get mapped into $|\det \, A_s|$ different solutions. Therefore, the error propagation from $\boldsymbol{\omega}$ to $\boldsymbol{v}$ is well-behaved if $A_s$ is non-singular, and it is legitimate to use the inverse matrix $A_s^{-1}$, as long as one remembers to apply it to all the vectors $\boldsymbol{\omega}^{\,c}$. 

Thus it is possible to give the mean and variance of the probability distribution function of $\boldsymbol{v}$, in a neighbourhood of each solution, in terms of the expectation value $\boldsymbol{\mu} := \E [\boldsymbol{\omega}]$ and covariance matrix $\Sigma_{a,b} := \text{Cov} (\omega_a, \omega_b) = \text{Cov} (\omega_a^{\,c}, \omega_b^{\,c})$:
\begin{align} 
  \E[\boldsymbol{v}^{\,c}]    &\ =\ A_s^{-1} \boldsymbol{\omega}^{\,c} \\
  \label{variance2}
  \text{Cov}(v_a^{\,c}, v_b^{\,c}) &\ =\ [A_s^{-1} \,\Sigma \,  (A_s^T)^{-1}]_{a,b}  \,,
\end{align}
in which the relation is exact, since the equations are linear. Notice that also the covariance matrix of $\boldsymbol{v}^{\,c}$ is independent of $c$. From now on we will omit the index $c$, since it will always be implied when giving a reconstruction of $\boldsymbol{v}$ depending on $\boldsymbol{\omega}$.\\

A characterization of error propagation can be given as follows. We have that $\boldsymbol{\omega} = \boldsymbol{w} + \delta \boldsymbol{w}$ represents the measured (perturbed) value of $\boldsymbol{w}$, with $\norm{\delta \boldsymbol{w}} \ll \norm{\boldsymbol{w}}$, where $\norm{\cdot}$ is any chosen vector norm; then the reconstructed solution of the system $A_s \boldsymbol{x} = \boldsymbol{w} + \delta \boldsymbol{w}$ is given by $\boldsymbol{x} \equiv v + \delta \boldsymbol{v}$. Then the following bound, arising from standard perturbation analysis, holds:
\begin{equation} \label{bound}
  \frac{\norm{\delta \boldsymbol{v}}}{\norm{\boldsymbol{v}}} \,\leq\, \norm{A_s} \norm{A_s^{-1}} \ \frac{\norm{\delta \boldsymbol{w}}}{\norm{\boldsymbol{w}}}
\end{equation}
where $\norm{A_s}$ is the operator norm induced by the chosen vector norm and $\kappa(A_s) := \norm{A_s} \norm{A_s^{-1}}$ denotes the condition number of matrix $A_s$ (which also depends upon the adopted vector norm). \\

Another measure of the magnitude of the errors that can be adopted is the \emph{uncertainty volume} in logarithmic variables:
\begin{equation} \label{vol}
  \text{Vol} (\boldsymbol{X}) := \sqrt{\text{det} \ \Sigma_{a b}} 
\end{equation}
where $\Sigma_{a b} := \text{Cov}(X_a,X_b)$ is the covariance matrix of set of random variables $\boldsymbol{X}$.

Hence, taking the square root of the determinant of~\eqref{variance2}, we get $\text{Vol}(\boldsymbol{v}^c) = \sqrt{\det(\Sigma)}/|\det{A_s}|$. That is, the uncertainty volume around each solution always shrinks by a factor $|\det \, A_s|$ 
\begin{equation}
  \frac{\text{Vol} (\boldsymbol{v}^c)}{\text{Vol} (\boldsymbol{\omega}^{\,c})} = \frac{1}{|\text{det} \ A_s|} \,.
\end{equation}
Notice, again, that the number of complex solutions is equal to $|\text{det} \, A_s|$, and that the above estimate gives the size of the error region around each valid solution. Thus there is a global conservation of uncertainty volume in our reconstruction method. However, not all complex solutions are physical; namely, as discussed in Sec.~\ref{sec:solving_methods}, we have to restrict ourselves to the case in which the solutions correspond to real values for the stray field intensities. Restricting to these physically relevant cases, the total uncertainty volume indeed can decrease, and it can never increase.

\subsection{Full error analysis} \label{sec:full_error_analysis}

\begin{figure*}
  \begin{ruledtabular}
  \begin{tabular}{ccc}
    & & \vspace*{-3mm}\\
    \includegraphics[height=.145\textheight]{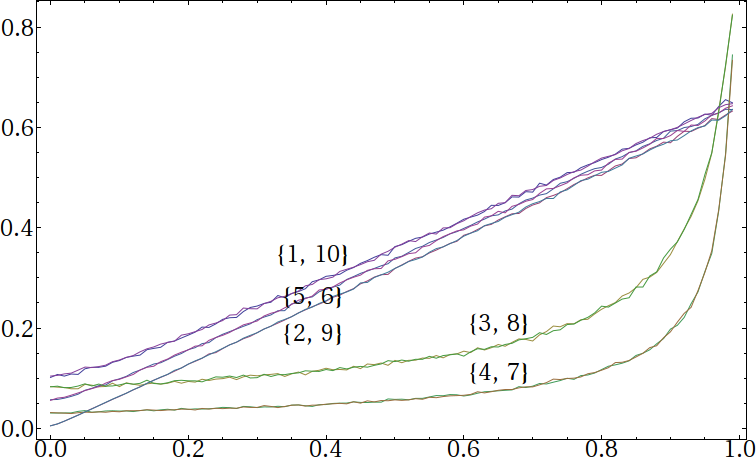} &
    \includegraphics[height=.145\textheight]{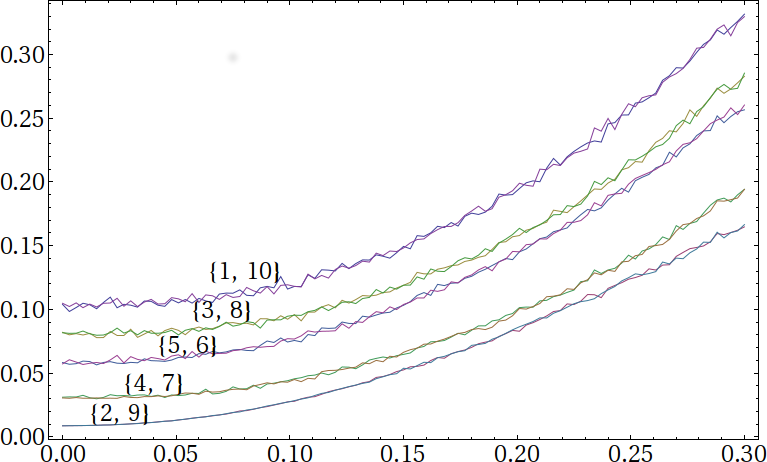} &
    \includegraphics[height=.145\textheight]{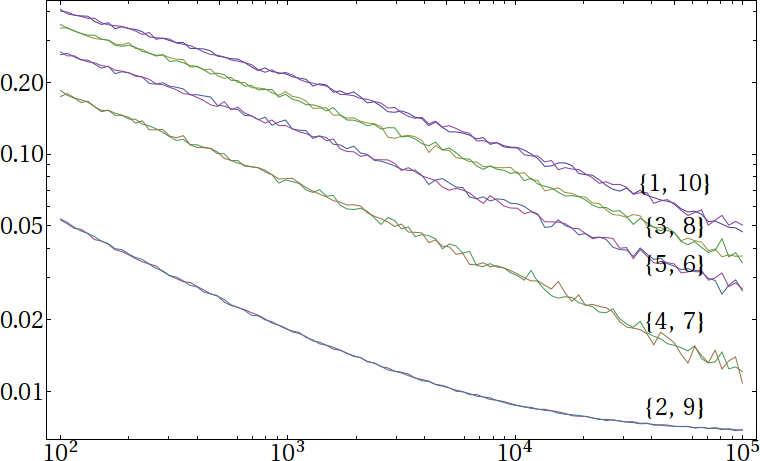} \\
    $X$-field, $q$-dependence &
    $X$-field, $\epsilon$-dependence &
    $X$-field, $M$-dependence
    \vspace*{3mm} \\    
    \includegraphics[height=.145\textheight]{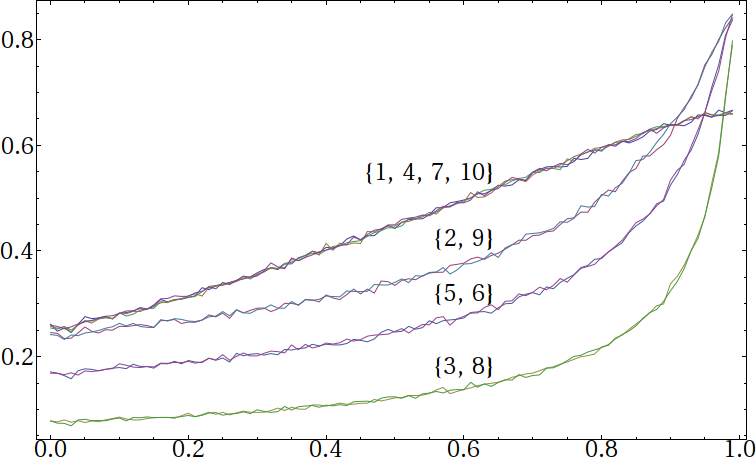} &
    \includegraphics[height=.145\textheight]{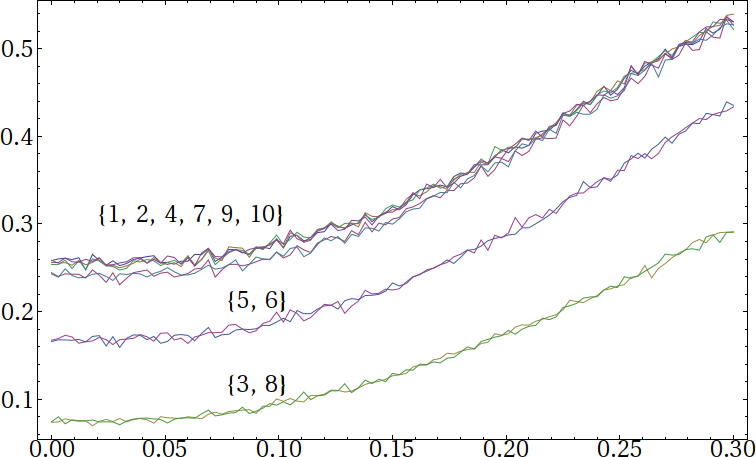} &
    \includegraphics[height=.145\textheight]{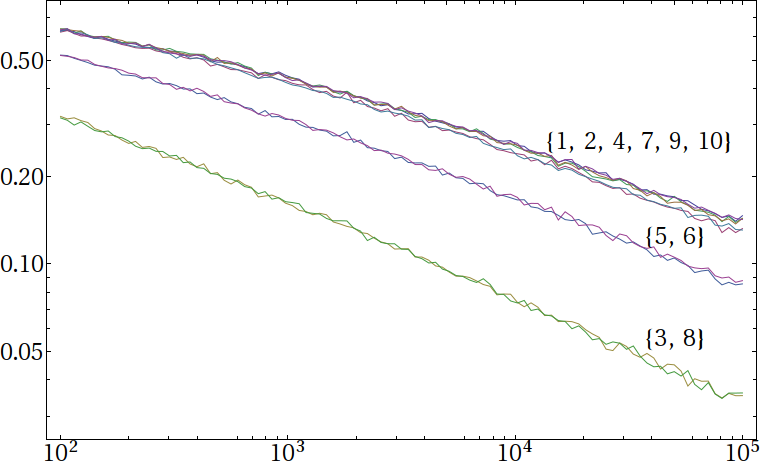} \\
    $Y$-field, $q$-dependence &
    $Y$-field, $\epsilon$-dependence &
    $Y$-field, $M$-dependence
    \vspace*{3mm} \\    
    \includegraphics[height=.145\textheight]{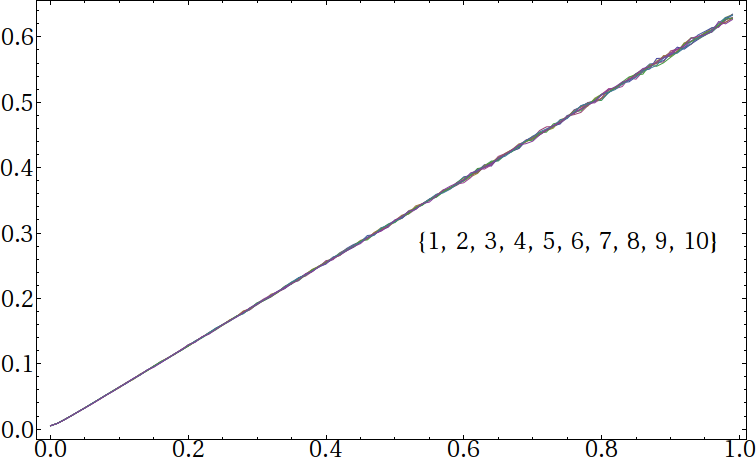} &
    \includegraphics[height=.145\textheight]{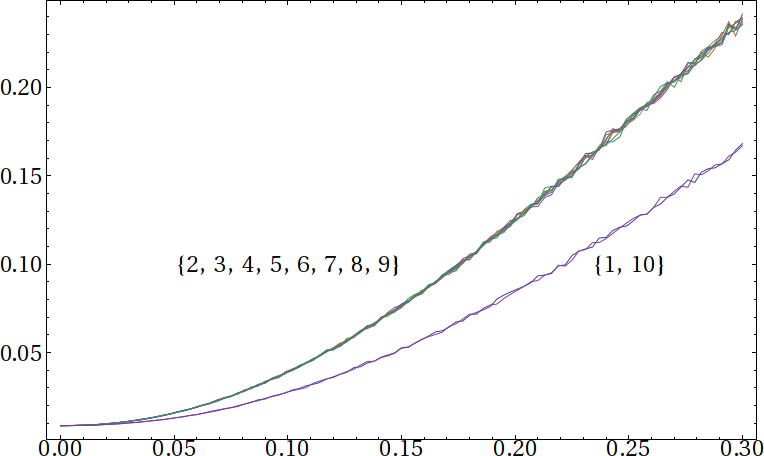} &
    \includegraphics[height=.145\textheight]{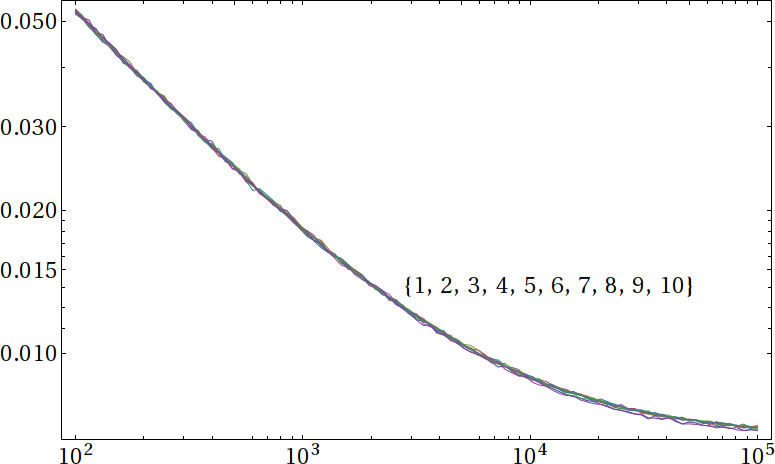} \\
    $Z$-field, $q$-dependence &
    $Z$-field, $\epsilon$-dependence &
    $Z$-field, $M$-dependence    
  \end{tabular}
  \end{ruledtabular}
  \caption{Summary of the effect on the \emph{reconstruction errors} $\mathcal{E}_a$ (see text) on all vertices $a \in V$ of an open-ended chain with $10$ vertices. The reconstruction errors are induced by the depolarizing channel (with parameter $q$), by the misalignment of the axis (with average misalignment given by $\epsilon$), and by the finiteness of the number of measurements of the correlators (given by $M$); we plot the results for the $X$-, $Y$-, and $Z$-field cases. In each plots we fix two of the parameters among $(q,\epsilon, M)$ and vary the third one; the reference values in all plots are $q=0.01$, $\epsilon =0.01$, $M=10^4$. The plots of the $M$-dependence of $\mathcal{E}_a$ are in bilogarithmic scale. Each curve in each plot represents the reconstruction error $\mathcal{E}_a$ for a specific vertex $a$ among the ten in the graph state, as indicated by the labels next to the corresponding curves. Each data point is obtained extracting an ensemble of $10^4$ field configurations, computing the reconstruction for each of these, and then averaging the reconstruction errors $\mathcal{E}_a$ over the ensemble. }
  \label{Fig2}
\end{figure*}

We will now deal with the full error analysis, in which the stray fields satisfy the constraints only approximately. That is, we analyse the case in which the fields are \emph{approximately} aligned to one of the three Cartesian axes $\hat{\boldsymbol{x}}$, $\hat{\boldsymbol{y}}$ or $\hat{\boldsymbol{z}}$. In all three situations we have to solve an approximate linear systems of the form
\begin{equation} \label{perturbed}
  A_s \boldsymbol{x} + \boldsymbol{f}_\epsilon (\boldsymbol{x}) \equal \boldsymbol{w} + \delta \boldsymbol{w}
\end{equation}
where $\boldsymbol{f}_\epsilon (\boldsymbol{x})$ is a vanishing function for $\epsilon \rightarrow 0$, with $\epsilon$ a global parameter which bounds the maximum misalignment of the rotation axis from one of the Cartesian axes. We call $\boldsymbol{x}$ a solution of this perturbed system, and with $\boldsymbol{v}$ a solution of the original unperturbed system. In Appendix~\ref{app:continuity} we show that, under some regularity assumptions, the magnitude of the perturbation in the solution is bounded by a continuous function of $\delta \boldsymbol{w}$ and of the rotation axis misalignments.

In principle, one should consider that in any realistic implementation there will be some incoherent noise acting alongside the local unitary rotations induced by the stray fields. Equivalently, we might consider the case in which the stray fields $\boldsymbol{\lambda}_a$ are not fixed vectors, but they vary in time. If these temporal variations are fast and stochastic, then it can be imagined that $\boldsymbol{\lambda}_a$ are random variables drawn according to some probability distribution.

The simplest model, but also the worst case scenario, for this source of incoherent noise is given by the uniformly depolarizing channel
\begin{equation} \label{depolarizing}
  \Phi_q (\rho) = (1-q) \rho + q \frac{\one}{2} \,,
\end{equation}
where $\rho$ is a single qubit density matrix and $0\leq q \leq 1$. It is not necessary to specify in which order the depolarizing channel and the local unitary rotations are applied, since they commute. From Eq.~\eqref{prob_diff_def} one immediately obtains the difference in probability between the measurement outcomes when both local unitary error sources and depolarizing noise $\Phi_q$ act on the system:
\begin{equation} \label{multiply_q}
  \DP_a^{\Phi_q} = (1-q) \,\DP_a
\end{equation} 
where $\DP_a$ is given by Eq.~\eqref{general}. This noise affects the results, favouring smaller values of $\beta_a$; for example, in the $Z$-field case all $\beta_a$ are just rescaled by a factor $(1-q)$. We point out that, for the $X$ or $Y$-field cases, for some of the qubits in the graph state the reconstruction algorithm may exhibit a natural resilience to the inaccuracy introduced by the uniformly depolarizing noise. In fact, from the expression in Eq.~\eqref{multiply_q} it can be seen that, to account for the action of the depolarizing noise, the differences in probabilities have to be multiplied by $(1-q)$ for all qubits. Thus the vector $\log(1-q) \, \boldsymbol{1} := \log(1-q) \, (1,1, \ldots, 1)^T$ has to be added to the known term in the linear equation~\eqref{linear_eq}, whose entries are the logarithm of the differences in probabilities. By linearity, the effect of the depolarizing noise on $\boldsymbol{w}$ (the vector of the estimators in logarithmic variables) is simply given by the addition of $\log(1-q) \times A_s^{-1} \boldsymbol{1}$. Some of the entries of the vector $A_s^{-1} \boldsymbol{1}$ might be zero, implying that the reconstruction of the fields for those entries is unaffected by the depolarizing noise.

We have tested through numerical simulation the performance of our reconstruction method when all the three noise factors considered above are present, i.e., the effect of the \emph{finiteness} of the number of measurements of the correlators, the \emph{misalignment} of the field from the promised direction, and the independent action of a \emph{depolarizing channel} on the qubits. These simulations provide evidence that, indeed, when all the noise factors are small (and the numbers of measurements of each correlator is large) then the accuracy in the reconstruction improves and, in principle, the inaccuracy can be brought arbitrarily close to zero.

We define the \emph{reconstruction error} $\mathcal{E}_a$ on any vertex $a \in V$ as
\begin{equation}
  \mathcal{E}_a := \left|\cos(\lambda_a) - \tilde{\beta}_a^{(reg)}\right| \,,
\end{equation}
in which $\cos(\lambda_a)$ is the true value of parameter and $\tilde{\beta}_a^{(reg)}$ is a regularization of the output $\tilde{\beta}_a$ of the reconstruction algorithm. The regularization consists in setting $\tilde{\beta}_a^{(reg)} = +1$ when $\tilde{\beta}_a \geq +1$ and $\tilde{\beta}_a^{(reg)} = -1$ when $\tilde{\beta}_a \leq -1$, so that the regularized value always lies in the interval $[-1,+1]$. An excerpt of the \emph{reconstruction errors} we have obtained, in the case of an open-ended chain with ten vertices, is given in Fig.~\ref{Fig2}.

A general trend that can be inferred from the numerical data is that -- since the correlation among the parameters is not taken into account -- the reconstruction error is minimal in the $Z$-field case, is larger in the $X$-field case, and is still larger in the $Y$-field case. This feature is expected, since the degree of the polynomial equations to be solved is minimal in the $Z$-field case, larger in the $X$-field case and still larger in the $Y$-field case. This implies that the cross-correlations between the estimators of the fields are maximal in the $Y$-field case and they are minimal in the $Z$-field case. 

Another feature that can be inferred from the data is that the precision of the estimation of the fields varies across the vertices of the graph and depends on the global connectivity of the graph itself. Notice that to produce the data shown in Fig.~\ref{Fig2} we have assumed that the model of the disturbance is the same for all ten vertices in the chain. Thus for symmetry reasons we expect that, averaging over all possible intensities of the stray field, for a open-ended chain with ten qubits the inaccuracy in the reconstruction of the fields on vertex $a$ and $11-a$ are the same.

Also, notice that the reconstruction error decreases approximately as $1/\sqrt{M}$ for small values of $M$, while for larger values of $M$ the reconstruction error approaches a non-zero asymptotic value, which is due to residual uncertainties introduced by the depolarizing noise and stray field misalignment.

Finally, we remark that at some of the vertices the reconstruction of the fields shows resilience to the depolarizing noise, for the reason given earlier. Explicitly, this happens for qubits on the vertices $3,4,7,8$ for the $X$-field case and for qubits on the vertices $2,3,5,6,8,9$ for the $Y$-field case, as can be easily checked computing $A_s^{-1} \boldsymbol{1}$ for these two cases. Consequently, it can be seen in Fig.~\ref{Fig2} that these qubits are almost insensitive to the magnitude $q$ of the depolarizing noise, provided that $q \lesssim 0.8$. When $q \rightarrow 1$, then the effect of the finiteness of the number of measurement $M$ renders unreliable the reconstruction of the fields on those qubits.

\section{Excursus: extension of the family of graph states} \label{sec:extension}

\begin{figure*}
  \includegraphics[scale=0.3]{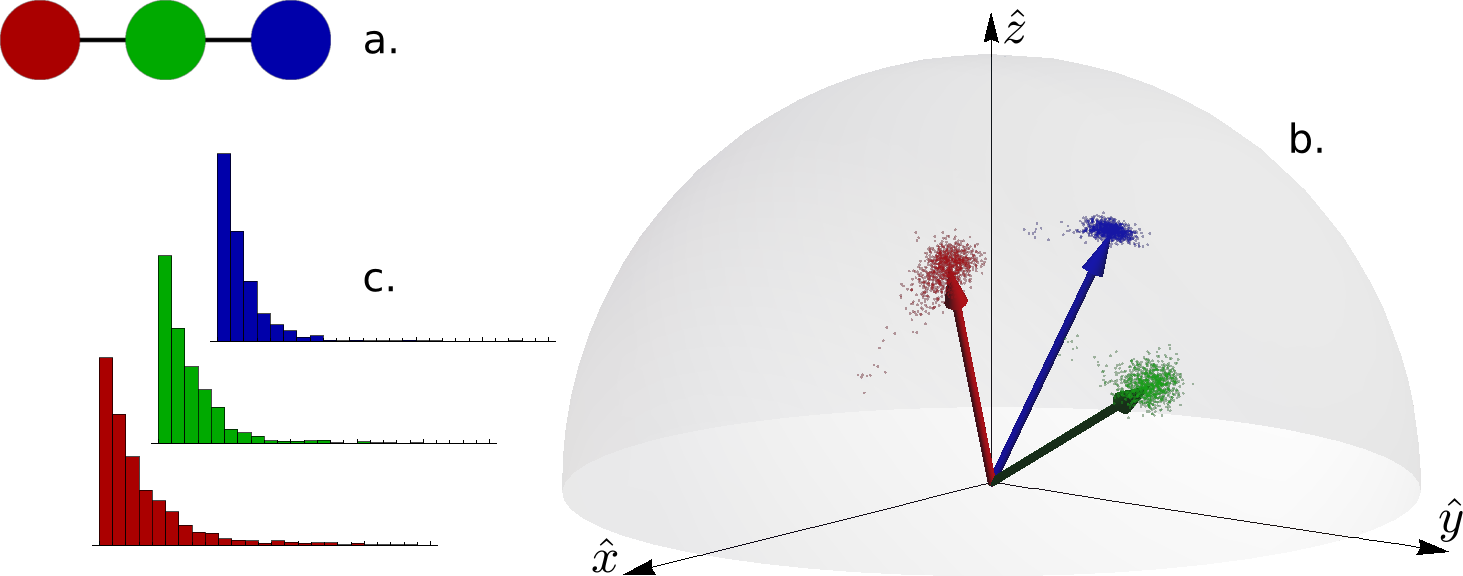}
  \caption{Example of the reconstruction of stray fields through the data collected from the stabilizer measurements associated to graph states with four different (randomly selected) logical Pauli bases. The graph state here is a linear chain with three qubits, as represented in Fig.~\textbf{a}; in these figures graphical objects sharing the same colour refer to the same qubit. The three vectors in Fig.~\textbf{b} point in the direction $\hat{\boldsymbol{n}}_a$ of the stray field at each vertex $a$ and have lengths proportional to $\beta_a = \cos \lambda_a$, where $\lambda_a = \gamma \Delta t |\boldsymbol{B}_a|$ is proportional to the strength of the field. We assume, without loss of generality, that the $\hat{\boldsymbol{z}}$ component of all vectors is positive, since in this representation opposite vectors are associated to the same physical stray field. We simulate the result of $10^4$ stabilizer measurements, for each of the three qubits in the graph state, for each of the four selected logical Pauli bases -- giving in total $3 \times 4 \times 10^4$ measurements of correlator operators. We then estimate from these $\cos \lambda_a$ and $\hat{\boldsymbol{n}}_a$ through the numerical minimization of cost function $C(\{\beta_a, \hat{\boldsymbol{n}}_a\}_{a \in V})$ given by Eq.~\eqref{cost}. We have repeated this procedure $10^3$ times; we display the resulting $10^3$ estimations of the parameters with coloured points in the hemisphere in Fig.~\textbf{b}. In Fig.~\textbf{c} the histograms show the number of points against the squared distance from the correct solution, i.e.\ the squared distance of the points from the tip of the corresponding arrow in plot \textbf{b}.}  
  \label{Fig1}
\end{figure*}

A completely different approach to estimating the stray fields consists in adopting different sets of stabilizer measurements in order to collect more information. In this way we do not need to assume that the fields are always aligned in a known direction. Using a different set of stabilizer measurements will produce a different graph state; but this is not an issue, as long as the functionality of this family of graph states is the same of that of the reference graph state.

Explicitly, we consider the following scenario. In the defining equation~\eqref{correlator}, the correlators are defined in terms of the product of Pauli-$X$ and Pauli-$Z$ matrices. Actually, many other correlators will yield graph states which have the same computational capabilities as the reference graph state. For example, we could as well use correlators in the form $Y_a \prod_{b\in\mathcal{N}_a}X_b$ (or in the form $Z_a \prod_{b\in\mathcal{N}_a}Y_b$). To see that these stabilizers define a graph state which is computationally equivalent to the original one it is enough to observe that these are obtained by a redefinition of the Pauli matrices, $(X,Y,Z)\rightarrow (Y,Z,X)$ (or $(X,Y,Z)\rightarrow (Z,X,Y)$), which is a transformation that preserves the commutation relations between the matrices -- and thus all the associated algebraic properties. Correspondingly, the local measurements necessary to perform MBQC have to be modified according to the new Pauli basis. 

These considerations apply to any transformation $(X,Y,Z)\rightarrow (\hat{\boldsymbol{r}}\cdot \boldsymbol{S},\, \hat{\boldsymbol{s}}\cdot \boldsymbol{S},\, \hat{\boldsymbol{t}}\cdot \boldsymbol{S})$, in which $\mathcal{B} \equiv (\hat{\boldsymbol{r}}, \hat{\boldsymbol{s}}, \hat{\boldsymbol{t}})$ is any right-oriented orthonormal basis of $\mathds{R}^3$. We call these \emph{logical Pauli matrices}. Correspondingly, Eq.~\eqref{general} becomes:
\begin{align} \label{rotated}
  \DP_a =  
    & \ \ \Big( (\hat{\boldsymbol{r}} \cdot \hat{\boldsymbol{n}}_a)^2 + \beta_a \big( 1 - (\hat{\boldsymbol{r}} \cdot \hat{\boldsymbol{n}}_a)^2 \big) \Big) \ \times \nonumber\\
    & \times \prod_{b \in \mathcal{N}_a}
      \Big( (\hat{\boldsymbol{t}} \cdot \hat{\boldsymbol{n}}_b)^2 + \beta_b \big( 1 - (\hat{\boldsymbol{t}} \cdot \hat{\boldsymbol{n}}_a)^2 \big) \Big) \,.
\end{align}
Thus, using three different graph states, associated to three different logical Pauli bases it is in principle possible to reconstruct all the parameters of the stray fields. Notice that in general there will be many physically acceptable solutions to a system of $3N$ polynomial equations with $3N$ unknowns. But then using $\ell \geq 4$ logical Pauli bases, from the corresponding graph states one obtains an overdetermined system of $\ell N$ equations with $3N$ unknowns, which almost certainly admits only one solution; moreover, having more independent equations usually provides a reconstruction which is more robust in presence of small uncertainties. Explicitly, we look for an approximate solution to the equations through numerical minimization of the following cost function:
\begin{align}\label{cost}
  & C \ \big(\{\beta_v, \hat{\boldsymbol{n}}_v\}_{v \in V}\big) := \nonumber\\
  & := \sum_{\substack{\mathcal{B} \ \in \\ \{\text{bases}\} }} 
      \sum_{a \in V} 
      \Big(\DP_a \big(\{\beta_b,\hat{\boldsymbol{n}}_b\}_{b \in V} \,; \, \mathcal{B}\big) - \DP_a^{(mes)}\big(\mathcal{B}\big)\Big)^2 \;.
\end{align}
Here $\DP_a (\{\beta_v,\hat{\boldsymbol{n}}_v\}; \mathcal{B})$ is computed from Eq.~\eqref{rotated} and it is a function of the field intensities $\beta_v$ and directions $\hat{\boldsymbol{n}}_v$ at the vertices in $\mathcal{N}_a' = \{ a \} \cup \mathcal{N}_a$, and of the choice of the logical Pauli basis~$\mathcal{B}$ (which can be identified with an orthonormal basis in $\mathds{R}^3$). With $\DP_a^{(mes)}(\mathcal{B})$ we denote the value for the probability difference estimated from syndrome measurements for vertex $a$ using the logical Pauli basis~$\mathcal{B}$. An example of the result that can be obtained via this method is reported in Fig.~\ref{Fig1}.

\section{Conclusions}

In this paper we have shown how to use the information collected from a set of stabilizer measurements in order to extract information about a unitary error process acting on a graph state. Since the stabilizer measurements are non-local operations, each of them depends on the errors that act on several different sites of the graph state. Therefore the measurement outcomes of these stabilizers are correlated.

We have studied in depth the case in which the local unitary errors are given by rotations aligned to one of the Cartesian axes. This condition is needed to make the problem solvable, i.e., we match the number of parameters to be reconstructed to the number of measured parameters; moreover this restriction allows us to find efficient reconstruction methods.

A surprising result arising from the fact that the stabilizer measurement outcomes are correlated is that in some circumstances it is not possible to reconstruct all the parameters of the unitary error -- even under the assumption that we can measure the stabilizers an arbitrary number of times. The possibility of completely reconstructing the error turns out to depend non-trivially on the connectivity structure of the graph state. Specifically, in order to determine the intensity of local unitary rotations around the $\hat{\boldsymbol{x}}$ axis, the adjacency matrix $A$ associated to a graph state $G$ has to be a full-rank matrix, whereas to reconstruct rotations around the $\hat{\boldsymbol{y}}$ axis the matrix $A + \one$ has to be full-rank, while rotations around the $\hat{\boldsymbol{z}}$ axis can always be reconstructed. 

We have ascertained for some classes of graph states whether it is possible to reconstruct all the errors. Hence we have studied the properties of the adjacency matrices of \emph{linear chains} (with open and closed boundary conditions), \emph{square lattices} in arbitrary number of dimensions, and GHZ states (these represented either as \emph{fully-connected} graphs or as \emph{star} graphs).

Finally we have shown that the performed reconstruction is resistant to various kinds of imperfections. In particular we have investigated the effect of a finite number of measurements of stabilizers, of the misalignment of the rotation axis and of the action of a depolarizing noise alongside the action of the unitary channel. In each of these cases, as long as the imperfections are small, the precision in the determination of the unitary channel is only mildly affected.

\section*{Acknowledgments}

We acknowledge support from the Austrian Science Fund (FWF) through the SFB FoQuS: F4012, and the Templeton World Charity Foundation grant TWCF0078/AB46.

\appendix

\section*{Appendices}

\section{Derivation of the error probability} \label{app:error_probability}

In this appendix we derive Eq.~\eqref{general}. This equation gives the difference between the probability of getting $\kappa_a=0$ and $\kappa_a=1$ when measuring the correlator $K_a = X_a \bigotimes_{b \in \mathcal{N}_a} Z_b$ on a graph state which is subject to local noise of the form 
$U_{field} = \prod_{a \in V}  U_a = \prod_{a \in V} e^{-i \lambda_a \hat{\boldsymbol{n}}_a \cdot \boldsymbol{S}_a /2}$.

We introduce the notation $\ket{\widetilde{G}} := U_{field} \ket{G}$; observe that $K_a$ has spectrum $\{+1,-1\}$, therefore there is a unitary mapping $V_a$ such that $K_a = {P_0}^a - {P_1}^a$, where ${P_j}^a$ with $j \in \{0,1\}$ is the projector on the $(-1)^j$-eigenspace of $K_a$. Hence we get that the difference in the probabilities is given by the expectation value of $K_a$ on $\ket{\widetilde{G}}$:
\begin{align}
 \DP_a & = \ \PP(\,\kappa_a = 0\ |\ \widetilde{G}\,) \; - \; \PP(\,\kappa_a = 1\ |\ \widetilde{G}\,) \nonumber\\
       & = \ \bra{\widetilde{G}}\; {P_0}^a  \;\ket{\widetilde{G}} 
           - \bra{\widetilde{G}}\; {P_1}^a  \;\ket{\widetilde{G}} \nonumber \\
       & = \ \bra{\widetilde{G}}\; ({P_0}^a - {P_1}^a)  \; \ket{\widetilde{G}} \nonumber \\
       & = \ \bra{\widetilde{G}}\; K_a \;\ket{\widetilde{G}} \,.
\end{align}
Now we can compute:
\begin{align}
  & K_a \, \ket{\widetilde{G}} = \nonumber\\
  & = K_a \ U_{noise} \ K_a \, \ket{G} \nonumber\\
  & = \left(X_a \bigotimes_{b \in \mathcal{N}_a} Z_b \right)
      \left[\prod_{c \in V} e^{-i \lambda_c \hat{\boldsymbol{n}}_c \cdot \boldsymbol{S}_c/2}\right]
      \left(X_a \bigotimes_{b \in \mathcal{N}_a} Z_b \right)
      \ket{G} \nonumber\\
  & = \left[\prod_{c \in (V \setminus \mathcal{N}_a' )} e^{-i \lambda_c \hat{\boldsymbol{n}}_c \cdot \boldsymbol{S}_c/2} \right] \cdot \nonumber\\
  & \ \cdot \left( 
             e^{-i \frac{\lambda_a}{2} \hat{\boldsymbol{n}}_a \cdot (X_a \boldsymbol{S}_a X_a)} 
           \prod_{b \in \mathcal{N}_a}
             e^{-i \frac{\lambda_b}{2} \hat{\boldsymbol{n}}_b \cdot (Z_b \boldsymbol{S}_b Z_b)} 
           \right)
           \ket{G}
\end{align}
where $\mathcal{N}_a' = \mathcal{N}_a \cup \{a\}$, and in which we have used that $K_a \ket{G} = \ket{G}$ and that $U e^{A} U^\dag = e^{UAU^\dag}$ when $U$ is a unitary operator. Hence:
\begin{align} \label{expectation}
  & \bra{\widetilde{G}}\; K_a \;\ket{\widetilde{G}} = \nonumber\\  
  & = \bra{G}\; U_{noise}^\dag \ K_a \ U_{noise}\ K_a \;\ket{G} \nonumber\\
  & = \bra{G}\
       e^{i \frac{\lambda_a}{2} \hat{\boldsymbol{n}}_a \cdot \boldsymbol{S}_a} \ \
       e^{-i\frac{\lambda_a}{2} \hat{\boldsymbol{n}}_a \cdot (X_a \boldsymbol{S}_a X_a)} \nonumber\\
  & \ \ \left( 
      \prod_{b \in \mathcal{N}_a}
       e^{i \frac{\lambda_b}{2} \hat{\boldsymbol{n}}_b \cdot \boldsymbol{S}_b} \ \
       e^{-i\frac{\lambda_b}{2} \hat{\boldsymbol{n}}_b \cdot (Z_b \boldsymbol{S}_b Z_b)} 
    \right)
    \ket{G} \,.
\end{align}

Actually, we only need to compute the part proportional to the identity in the operator in the last line of Eq.~\eqref{expectation}, since $\bra{G}X\ket{G} = \bra{G}Y\ket{G} = \bra{G}XZ\ket{G} = 0$ for any non-trivial graph $G$, i.e. for graphs in which there is no vertex which is completely disconnected from all other vertices (no isolated vertices). Indeed, for the Pauli $Z$ operation on the vertex $a$ we have:
\begin{align}
\bra{G} \ Z_a \ \ket{G} 
  & = \bra{G} \ Z_a \, K_a \ \ket{G} = - \bra{G} \ K_a \, Z_a \ \ket{G} \nonumber \\
  & = - \bra{G} \ Z_a \ \ket{G} = 0 \,,
\end{align}
in which we have used twice $K_a \ket{G} = \ket{G}$ and the anti-commutation relation between $Z_a$ and $K_a$. An analogous procedure shows that $\bra{G} Y_a \ket{G} = 0$, since also $Y_a$ and $K_a$ anti-commute. Then, for the Pauli $X$ operation on vertex $a$ we have (provided that $a$ has at least one neighbour):
\begin{align}
\bra{G}\; X_a\; \ket{G} = 
    \bra{G}\; X_a K_a \; \ket{G} = \bra{G} \prod_{b \in \mathcal{N}_a} \!\! Z_b \; \ket{G} = 0 \,.
\end{align}

Finally, we can evaluate the part proportional to the identity in Eq.~\eqref{expectation} applying the exponentiation formula for Pauli matrices:
\begin{equation}
  e^{i \lambda (\hat{\boldsymbol{n}} \cdot \boldsymbol{S}) } = \one \cos(\lambda) + i\,(\hat{\boldsymbol{n}} \cdot \boldsymbol{S}) \sin(\lambda)
\end{equation}
together with the identity:
\begin{equation}
  S_\alpha\,S_\beta\,S_\alpha = -(-1)^{\delta_{\alpha,\beta}} S_\beta \,,
\end{equation}
for $\alpha, \beta \in \{x,y,z\}$, obtaining:
\begin{align} \label{expectation2}
  & \bra{\widetilde{G}}\; K_a \;\ket{\widetilde{G}} = \nonumber\\  
  & = \bra{G}\ \one_a 
      \left[ 
        \cos^2 \frac{\lambda_a}{2} +
        \left( (n_{x}^a)^2 - (n_{y}^a)^2 - (n_{z}^a)^2\right)
        \sin^2 \frac{\lambda_a}{2}
      \right]\nonumber\\
  &\ \ \,\prod_{b \in \mathcal{N}_a} \one_b
      \left[ 
        \cos^2 \frac{\lambda_b}{2} + 
        \left( (n_{z}^b)^2 - (n_{x}^b)^2 - (n_{y}^b)^2 \right)
        \sin^2 \frac{\lambda_b}{2}
      \right]
    \ket{G} \nonumber\\
  & = \ \left[ 
        (n_{x}^a)^2 +
        \left( 1 - (n_{x}^a)^2 \right)
        \cos \lambda_a
      \right]\nonumber\\
  & \ \ \prod_{b \in \mathcal{N}_a}
      \left[ 
        (n_{z}^b)^2 + 
        \left( 1 - (n_{z}^b)^2 \right)
        \cos \lambda_b
      \right] \,.
\end{align}
This is exactly the expression given in Eq.~\eqref{general}, once one writes $\beta_a := \cos \lambda_a$.

\section{Continuity of the reconstruction in presence of small misalignments} \label{app:continuity}

Suppose that the rotation axes of the rotation generated by the stray field is almost parallel to $\hat{\boldsymbol{x}}$, $\hat{\boldsymbol{y}}$ or $\hat{\boldsymbol{z}}$, but is slightly misaligned. In these three cases we can rewrite the parameters $n_x^a$ and $n_z^a$ (which together define the direction in which the field points for each $a \in V$) as follows:
\begin{align}
  & (n_x^a)^2 =1-\epsilon_a \quad (n_z^a)^2 = \tilde{\epsilon}_a\white{-1} \qquad X\text{-field} \nonumber\\
  & (n_x^a)^2 =\epsilon_a\white{-1} \quad (n_z^a)^2 =\tilde{\epsilon}_a\white{-1} \qquad Y\text{-field}\nonumber\\
  & (n_x^a)^2 =\epsilon_a\white{-1} \quad (n_z^a)^2 = 1-\tilde{\epsilon}_a \qquad Z\text{-field}
\end{align}
where each $\epsilon_a$ and $\tilde{\epsilon}_a$ is assumed to be small. Then we write Eq.~\eqref{general} for these three cases with the probabilities $\DP$ replaced by the measured rate differences $\Delta R$, we take the logarithm of both sides, we expand $\ln(1+\epsilon) = \epsilon + \mathcal{O}(\epsilon^2)$ assuming that $\epsilon_a / \beta_a \ll 1$ and $\tilde{\epsilon}_b / \beta_b \ll 1$, finally obtaining:
\begin{align} \label{perturbed_explicit}
    \ln \Delta R_a  
  & \equal \ \sum_{b \in \mathcal{N}_a} \ln \beta_b \ + \nonumber \\
  & + \ \left[\epsilon_a(\beta_a-1) + \sum_{b \in N_a} \tilde{\epsilon}_b\frac{1-\beta_b}{\beta_b} \right] 
    + \mathcal{O}(\epsilon^2)\,,\nonumber\\
    \ln \Delta R_a  
  & \equal \ \ln \beta_a + \sum_{b \in \mathcal{N}_a} \ln \beta_b \ + \nonumber \\
  & + \ \left[\epsilon_a \frac{1-\beta_a}{\beta_a} + \sum_{b \in N_a} \tilde{\epsilon}_b\frac{1-\beta_b}{\beta_b} \right] 
    + \mathcal{O}(\epsilon^2)\,,\nonumber\\
    \ln \Delta R_a  
  & \equal \ \ln \beta_a \ + \nonumber \\
  & + \ \left[\epsilon_a\frac{1-\beta_a}{\beta_a} + \sum_{b \in N_a} \tilde{\epsilon}_b (\beta_b-1) \right]
    + \mathcal{O}(\epsilon^2)\,,
\end{align}
for the $X$-field, $Y$-field and $Z$-field case respectively. In all three cases we need to solve an approximate linear systems in the form
\begin{equation} \label{perturbed_app}
  A_s \boldsymbol{x} + \boldsymbol{f}_\epsilon (\boldsymbol{x}) \equal \boldsymbol{w} + \delta \boldsymbol{w}
\end{equation}
where we use the notation $\boldsymbol{w} = {(\ln \DP_1, \ldots , \ln \DP_N)}^T$, $\boldsymbol{\omega} = {\boldsymbol{w}+\delta\boldsymbol{w}} = {(\ln \Delta R_1, \ldots , \ln \Delta R_N)}^T$ and where $\boldsymbol{x} \equiv {\boldsymbol{v} + \delta \boldsymbol{v}} = {(\ln \beta_1, \ldots , \ln \beta_N)}^T$ is the solution of the perturbed system of equations~\eqref{perturbed_app}, being $\boldsymbol{v}$ a solution of $A_s \boldsymbol{v} \equal \boldsymbol{w}$, i.e.\ a solution of the original unperturbed system. Furthermore, we have introduced a global scale parameter $\epsilon$ given by $\epsilon := \underset{a \in V}{\max}\,\max\,(\epsilon_a, \tilde{\epsilon}_a)$.

This perturbed (nonlinear) system can in principle be solved recursively, through the following relations:
\begin{equation} \label{recursion}
\begin{cases}
  \boldsymbol{x}_0\quad\text{ s.t.}& A_s\,\boldsymbol{x}_0\quad = \boldsymbol{\omega} \\
  \boldsymbol{x}_{j+1} \text{ s.t.}& A_s\,\boldsymbol{x}_{j+1}  = \boldsymbol{\omega} - \boldsymbol{f}_\epsilon (\boldsymbol{x}_j) 
\end{cases}
\end{equation}
where we use the equality symbol $=$ rather then $\equal$ as we will focus on a particular solution of the system~\eqref{perturbed_app}.  We have to show that the solution of the perturbed system varies continuously with respect to the input data and field directions, i.e.\ $\delta \boldsymbol{v} \equiv (\boldsymbol{x} - \boldsymbol{v}) \rightarrow 0$ when $\delta \boldsymbol{w} \rightarrow 0$ and $\epsilon \rightarrow 0$.

Initially we have to check that the vector $\boldsymbol{v}$, obtained from solving the unperturbed linear system, satisfies the constraints
\begin{align}
  & \exists \ r, R > 0 \ \text{ such that :} \nonumber\\
  & \qquad r \leq \norm{\boldsymbol{v}}_\infty \leq R  \,, \\
  & \forall \ a \in V: \nonumber\\
  & \qquad \Re (v_a) < 0 \,,
\end{align}
where $\norm{v}_\infty := \underset{a \in V}{\max}\,|v_a|$. Then it is straightforward to verify that for $\boldsymbol{f}_\epsilon (\boldsymbol{v})$ given by any of the three expressions in square brackets in Eq.~\eqref{perturbed_explicit}, the following bound holds:
\begin{equation}
  \norm{\boldsymbol{f}_\epsilon (\boldsymbol{v})}_\infty \leq 2\epsilon \, (d+1) \, e^R
\end{equation} 
where $d := \underset{a \in V}{\max}\,|\mathcal{N}_a|$, i.e.\ $d$ is the maximum number of connections for any vertex in the graph. Also, using the triangle inequality, it can be verified that:
\begin{align}
  & r_0 \leq \norm{\boldsymbol{x}_0}_{\infty} \leq R_0 \\
  \text{with}\quad & r_0\;=\;r - \norm{A_s^{-1}}_\infty \norm{\delta \boldsymbol{w}}_\infty\,,\nonumber\\ 
                   & R_0  =  R + \norm{A_s^{-1}}_\infty \norm{\delta \boldsymbol{w}}_\infty\,,\nonumber
\end{align} 
and, provided that $\norm{\delta \boldsymbol{w}}_\infty < r / \norm{A_s^{-1}}_\infty$, we also have $r_0 >0$. Then the recursive relation~\eqref{recursion} allows us to bound the norm of the $j$-th iterative solution $\boldsymbol{x}_j$, provided that
\begin{equation} \label{lambert_condition}
  R_0 < W \left( \frac{r_0}{\norm{A_s^{-1}}_\infty \, 2 \epsilon \, (d+1) \, e }\right) \,,
\end{equation}
where $W(\cdot)$ is the Lambert $W$-function, i.e.\ $W(z)$ is the solution of $x\,e^x = z$. This can be shown bounding $\norm{\boldsymbol{x}_j}_\infty$ with $\norm{\boldsymbol{x}_j}_\infty < R_j$, in which again $R_j$ is given through a recurrence relation:
\begin{align}
  \norm{\boldsymbol{x}_{j+1}}_\infty 
    & \leq R_0 \left( \frac{\norm{\boldsymbol{x}_{j+1}}_\infty}{\norm{\boldsymbol{x}_0}_\infty} \right) \nonumber\\
    & \leq R_0 \left( 1+ \frac{\norm{A_s^{-1}}_\infty}{r_0} \, 2 \epsilon \, (d+1) \, e^{R_j} \right) \nonumber\\
    & \equiv R_{j+1}
\end{align}
and we obtain $R_\infty$ through the relation $R_{j+1} = R_j$; this can be solved when Eq.~\eqref{lambert_condition} is satisfied, and if this is the case we obtain:
\begin{equation} \label{R_bound}
  R_\infty = R_0 - W(-C \, R_0 \, e^{R_0})
\end{equation}
with $C := \frac{\norm{A_s^{-1}}_\infty}{r_0} \, 2 \epsilon \, (d+1)$. Notice that $R_\infty \rightarrow 0$ when $R_0 \rightarrow 0$, which can be enforced if $\epsilon \rightarrow 0$ and $\delta \boldsymbol{w} \rightarrow 0$. 

In conclusion we have shown that the perturbed solution $\boldsymbol{x}$ (which has to exist for physical consistency) has a distance from the unperturbed solution $\boldsymbol{v}$ which is at most $R_\infty$, and that $R_\infty$ goes to zero when both the magnitude of the misalignment and the inaccuracy in measuring the rates go to zero.


\end{document}